\documentclass[12pt,draft]{article}
 
\usepackage{amsmath}
\usepackage{amsthm}
\usepackage{amssymb}
\usepackage{epsf}
\usepackage{rotating}
\usepackage{wrapfig}

\def\drawbox#1#2{\hrule height#2pt
        \hbox{\vrule width#2pt height#1pt \kern#1pt
              \vrule width#2pt}
              \hrule height#2pt}
\def\Fund#1#2{\vcenter{\vbox{\drawbox{#1}{#2}}}}
\def\Asym#1#2{\vcenter{\vbox{\drawbox{#1}{#2}
              \kern-#2pt       % line up boxes
              \drawbox{#1}{#2}}}}
\def\fund{\Fund{6.5}{0.4}}

\def\href#1#2{#2}

\begin{document}

\thispagestyle{empty}
\rightline{CALT-68-2214}
\rightline{HUB-EP-99/08}
\rightline{MIT-CTP-2833}
\rightline{hep-th/9903093}
\vspace{0.6truecm}
\centerline{\bf \Large Mirror Symmetries for Brane Configurations}
\vspace{.4truecm}
\centerline{\bf \Large and
Branes at Singularities}

\vspace{0.6truecm}
\centerline{\bf  Mina Aganagic
%\footnote{mina@theory.caltech.edu}
}
\vspace{.2truecm}
{\em \centerline{California Institute of Technology, Pasadena CA 91125, USA}}
{\em\centerline{mina@theory.caltech.edu}}
\vspace{.4truecm}
\centerline{\bf 
Andreas Karch
%\footnote{karch@ctp.mit.edu}
}
\vspace{.2truecm}
{\em \centerline{Center for Theoretical Physics, MIT, 
Cambridge, MA 02139, USA}} 
{\em\centerline{karch@ctp.mit.edu}}            
\vspace{.4truecm}
\centerline{\bf Dieter L\"ust
%\footnote{luest@physik.hu-berlin.de}
\ and \
Andr\'e Miemiec
%\footnote{miemiec@physik.hu-berlin.de}
}

\vspace{.2truecm}
{\em 
\centerline{Humboldt-Universit\"at, Institut f\"ur Physik,
D-10115 Berlin, Germany}}
{\em \centerline{luest,miemiec@physik.hu-berlin.de}}

\vspace{0.5truecm}
%%%%%%%%%%%%%%%%%%%%%%%%%%%%%%%%%%%%%%%%%%%%%%%%%%%%%%%%%
%\vspace{.4truecm}
\begin{abstract}
We study local mirror symmetry 
on non-compact Calabi-Yau manifolds (conifold type of singularities)
in the presence of D3 brane probes. Using
an intermediate brane setup of NS 5-branes
`probed' by D4 resp. D5 branes, we can explicitly T-dualize three
isometry directions to relate a 
non-compact Calabi-Yau manifold to its local mirror. The
intermediate brane setup is the one that is best suited to read off the
gauge theory on the probe. Both intervals and boxes 
of NS 5-branes appear as brane setups.
Going from one to the other is equivalent to
performing a conifold transition in the dual geometry. One result of
our investigation is that the brane box rules as they have been discussed so
far should be generalized. Our new rules do not need diagonal fields localized
at the intersection. The old rules reappear on baryonic branches of
the theory. 
\end{abstract}

%\bigskip \bigskip
\newpage

\tableofcontents\newpage

%%%%%%%%%%%%%%%%%%%%%%%%%%%%%%%%%%%%%%%%%%%%%%%%%%%%%%%%%%%%%
%       SECTION:  Introduction
%%%%%%%%%%%%%%%%%%%%%%%%%%%%%%%%%%%%%%%%%%%%%%%%%%%%%%%%%%%%%

\font\markfont=cmbxsl10 scaled \magstep1
\font\bigmarkfont=cmbxsl10 scaled \magstep3

\section{Introduction}
\label{sectionintroduction}

Mirror symmetry is 
a symmetry which relates topologically distinct pairs of
(complex) $d$-dimensional
Calabi-Yau manifolds to each other \cite{mirrorreview1,mirrorreview2}.
In the past, mirror symmetry was mainly discussed for compact Calabi-Yau
manifolds. If ${\cal M}$ and ${\cal W}$ constitute a Calabi-Yau 
mirror pair, it follows that
\begin{equation}
h^{1,1}({\cal M})=h^{1,d-1}({\cal W}), \quad h^{1,d-1}({\cal M})=
h^{1,1}({\cal W}),
\end{equation}
where $h^{p,q}$ denotes the dimension of the cohomology of p-holomorphic and
q-anti-holomorphic forms.
This observation leads to very powerful predictions,
namely it identifies the classical moduli space of complex structures
of ${\cal M}$ (${\cal W}$) with the quantum moduli space
of the complexified K\"ahler classes of ${\cal W}$ (${\cal M}$), which includes quantum corrections from holomorphic curves.

The Calabi-Yau mirror symmetry was originally discovered  in the
context of perturbative string compactifications
on Calabi-Yau three-folds, where the mirror operation just corresponds
to a sign flip of the charges  of the $U(1)$ currents in the
underlying superconformal $n=2$ algebra \cite{mirror1,mirror2}.
It implies that the perturbative heterotic string is invariant under
the mirror symmetry, whereas the perturbative
type IIA and IIB superstrings on Calabi-Yau three-folds are mapped
onto each other by the mirror operation.
More recently, assuming  
that mirror symmetry extends 
to a symmetry of the full non-perturbative string theory, 
authors of \cite{SYZ} provided a geometric interpretation
to the mirror map. They showed that ${\cal M}$ has a quantum mirror 
provided it is a $T^d$ fibration
over a d-dimensional base $B$, where the fibers $T^d$ are 
Lagrangian submanifolds relative to the K\"ahler form.
Mirror symmetry is than the $T$-duality transformation
with respect to the volume of $T^d$, i.e. it inverts all radii of $T^d$.

In this paper we discuss T-duality and mirror symmetry for type II string
theory on a particular class of Calabi-Yau
spaces.
%\footnote{ Some aspects of T-duality in relation of non-compact 
%                Calabi-Yau spaces were discussed in \cite{KKL}.}.
In our point of view, the geometry will serve as a background,
and we will study the gauge theories living on 
D-branes probing the manifolds.

Consider first a D3 brane probing a Calabi-Yau manifold ${\cal M}$.
At a smooth point in  ${\cal M}$ the tangent space is ${\mathbb R}^6$
and the D3 brane will have $N=4$ supersymmetry on the world volume.
To get something more interesting, we have to consider Calabi-Yau
manifolds with singularities. 
Since we are only interested in the local physics near the singularity, 
our manifolds will all be non-compact Calabi-Yau spaces which, if
one desires to, can be viewed as having a completion to compact 
Calabi-Yau manifold. 

Some of the Calabi Yau manifolds we will study have hyperquotient singularities
that can be obtained as orbifolds of the well known conifold singularity
${\cal C}$, and so are of the form ${\cal C}/\Gamma$, where $\Gamma$ is
a discrete symmetry group a conifold admits. 
Recently the gauge theory of D3 brane probing
a conifold singularity was derived in \cite{klebwit}.
The theory of D3 brane on ${\cal C}/\Gamma$ is then defined as
a quotient of the theory on ${\cal C}$ by $\Gamma$.\footnote{The discussion of 
D3 branes
on 6-dimensional singularities is  very closely 
related to the  by now famous AdS/CFT
correspondence. Superconformal $N=2$ or $N=1$ (and also $N=0$) gauge
theories can be constructed as the duals of 
supergravity on $AdS_5\times X^5$,
where $X^5$ is a certain five-dimensional (Einstein) manifold. First, for the
case of D3 branes on six-dimensional orbifold singularities
${\cal O}={\mathbb R}^6/\Gamma$, where $\Gamma$
is a discrete group, $X^5$ is given by $S^5/\Gamma$, as discussed
in \cite{KachruFinite,Lawrence}. 
The corresponding orbifold gauge theory can 
be calculated using string perturbation theory.
The conifold singularities were later obtained in \cite{klebwit}, where for
the simplest conifold the corresponding Einstein space $X^5$ 
is the homogeneous space 
$T^{1,1}=(SU(2)\times SU(2))/U(1)$.
Further conifold type of singularities 
were recently discussed in \cite{Uranga}.}

We actually need to be a little bit more precise about the meaning of 
singularities in string theory. 
The Calabi-Yau singularities often have topologically distinct resolutions,
the conifold singularity being the simplest example.
In that case, we could either deform the 
complex structure of the conifold or its 
K\"ahler structure to obtain a
smooth manifold. Now, when we discuss the conifold in string theory 
we have to specify the means of smoothing the singularity. 
This is because
on the K\"ahler side of the conifold it suffices to turn on the $B$ field 
flux to obtain theory isomorphic to that on the smooth space. 
The D3 brane theory constructed in \cite{klebwit}
is the theory on the resolution of the conifold.
Taking a quotient of this theory by $\Gamma$ the 
resulting theories should be viewed as coming from
the K\"ahler side of the singularity. 

Locally, complex and K\"ahler structure moduli spaces decouple.
Thus, if we are interested in a neighborhood in which 
${\cal M}$ develops a singularity through degeneration of
its K\"ahler structure, we can take the complex structure to be nice and 
smooth, and therefore trivial. Canonically mirror symmetry acts by
exchange of complex and 
K\"ahler structure.  If $({\cal M}, {\cal W})$ form a mirror pair,
it is the complex structure of ${\cal W}$ that will be interesting.

The mirror geometries of the singularities will be constructed 
precisely in the spirit of \cite{SYZ}, namely
by performing T-duality transformations
around three isometric directions of the geometric singularity.
The 
singularities we are interested in
have a toric description so we can equivalently
\cite{leungvafa} apply the local mirror map
in the toric language \cite{KMV}.
The first point of view will be more useful for us, since it
will allow us to follow the action of mirror symmetry on D-branes.
Since the mirror symmetry acts in the space transverse to the D3
branes, the IIB gauge theory of D3 branes probing the space ${\cal M}$ will
be mirror to an identical gauge theory but now due to IIA D6 branes wrapping a 3-cycles in ${\cal W}$. ``Mirror'' of the D3 brane at a smooth point will be  
a D6 brane wrapping $T^3$. What will be the mirror of a D3 brane
at the singular point? The mirror D6 brane will wrap a 
three cycle which is still a special Lagrangian, but is now a  
degenerate three cycle which is homologous to the fiber at a generic point.

As it is known already for some time \cite{dualNS,DonD}, 
the geometric orbifold or
conifold singularities are T-dual to a certain
number of Neveu-Schwarz (NS) 5-branes.
This T-duality can be used 
\cite{smith,haur,ACL,Uranga,dasguptamukhi,lopez} 
to transform the D3 branes probing a 
singularity into a pure brane configuration of intersecting NS branes
and D branes of the Hanany-Witten type  \cite{HW,hanzaf}. 
It is this fact that we will systematically explore here.

In our case manifold ${\cal M}$ has three isometries
on which T-duality $T_{\rm mirror}$ can be performed to obtain ${\cal W}$.  
We can  write the mirror transform $T_{\rm mirror}$
as a composition of two dualities
$T_U$ and $T_V$, such that
starting with 
a singularity ${\cal M}$ and acting
with $T_{\rm mirror}$ on that space, we will first dualize to a certain brane
configuration and subsequently further  to the mirror geometry 
${\cal W}$. 
From the brane point of view (taking NS5 branes to have $x^{0,1,2,3}$
as common directions and extend along $x^{4,5}$ and $x^{8,9}$)
so we will call $T_U = T_6$ and $T_V =T_{48}$. 
From these two differently oriented NS branes we can build boxes or intervals,
respectively, each giving rise to a pair of $(T_U,T_V)$ dual
mirror geometries. As is well known, one can suspend D4 branes
on the intervals, and D5 branes on the boxes to obtain four dimensional gauge
theories on the D brane world volumes. 
T-duality will map these to either probe D3
branes or the D6 branes wrapping three-cycles of the mirror geometry.  
The resulting field theory should be the same in all the T-dual realizations.

Using these relations we can derive the rules that govern
which gauge theory is encoded in a given brane setup.
%In this way we will be able to show how the K\"ahler (complex structure)
%parameters of ${\cal M}$ are mapped to the complex structure
%(K\"ahler) parameters of ${\cal W}$.

%Of course we could also take as the starting point one of the 
%brane configurations; the mirror transformation will then map
%this brane configuration to the corresponding mirror branes, where
%the intermediate step is now given by a geometric singularity.
%Specifically this would imply that the brane boxes and the brane interval
%theories are mirror to each other.
%Lifting these two types of brane configurations to M-theory,
%the interval theories are described by supersymmetric 2-cycles embedded
%into ${\mathbb R}^6$ \cite{witten4d}, 
%whereas the M-theory lift of the brane bo
%models is characterized by a supersymmetric 3-cycle in ${\mathbb R}^6$
%\cite{miemiecetal}.
%So the mirror transformation should map every supersymmetric 2-cycle
%to a supersymmetric 3-cycle and vice versa.
%It actually turns out that only very simple singularities have both
%a interval as well as a box dual.

The paper is organized as follows. In the next section  we will introduce
the relevant geometries, namely the conifold singularities and
the orbifold singularities and their generalizations. 
In the third section we will discuss the gauge theories that
appear on the D3 brane probes. 
In section four we will introduce the T-dual
brane setups -- T-duality by T$_U$
or T$_V$ respectively -- and will discuss the T-duality without the probe.
Putting together the two T-duality transformations
we will see the mirror geometries emerging.
In section five we than incorporate the D3 brane probes. We will
find that the brane box is the natural dual of the blowup
of the orbifolded conifold and of the deformed generalized conifold.
In order to incorporate this result we have to modify the brane
box rules of Hanany and Zaffaroni \cite{hanzaf}. 
Their gauge theories reappear in a special corner of moduli
space.
Our new construction makes
some aspects of the box rules more transparent.
In section six we will wrap up by considering some related issues. We will show
that by the same transformation T$_{468}$ mirror symmetry can be defined
for brane setups as well, turning 2-cycles into 3-cycles. 
We will show how to put both,
the box
and the interval together in one picture.
This way we obtain a
domain wall in an $N=1$ 4d gauge theory that
lifts to M-theory via a $G_2$ 3-cycle as in \cite{witteng2}.

%%%%%%%%%%%%%%%%%%%%%%%%%%%%%%%%%%%%%%%%%%%%%%%%%%%%%%%%%%%%%
%       SECTION:  GEOMETIES
%%%%%%%%%%%%%%%%%%%%%%%%%%%%%%%%%%%%%%%%%%%%%%%%%%%%%%%%%%%%%

\section{The geometries}
\label{sectiongeometries}

\subsection{Conifold}

The simplest isolated singularity a three dimensional Calabi Yau manifold 
can develop is the conifold:
\begin{equation}
   \label{con} {\cal C}:\quad xy-uv=0
\end{equation}
The singularity is located at $x=y=u=v=0$ where the manifold fails
to be transverse: $f=xy-uv=0$, $\partial_i f =0$ have a common 
solution there.
There are two ways of smoothing the singularity, resulting in topologically 
distinct spaces.
\begin{itemize}
\item
   The so called small resolution -- replacing the singular point
   by a ${\mathbb{CP}}^1$, thereby changing the K\"ahler structure.
   The resulting space has \hbox{$h^{1,1} = 1$}, \hbox{$h^{2,1} = 0$}. 
\item  
   By deformation of the defining equation, thereby changing the complex 
   structure. After the deformation, $h^{1,1} = 0$, $h^{2,1} = 1$. 
\end{itemize}
%The resulting spaces have Hodge numbers reversed. The blowup and 
%deformation of the conifold is in fact our first example of mirror symmetry.

\subsubsection*{Small Resolution}

There are many ways in which one can exhibit the small resolution of 
the conifold. The one particularly well suited for our purposes is 
as follows. One can solve equation (\ref{con}) by simply putting 
\begin{equation}
\label{con2}\quad x=A_1 B_1 , \quad y =A_2 B_2, \quad 
u=A_1B_2,  \quad v = A_2 B_1,
\end{equation}
where $A_i, B_j \in {\mathbb C}^4$.
%%%%%%%%%%%%%%%%%%%%%%%%%%%%%%%%%%%%%%%%%%%%%%%%%%%%%
%   BILD: CONIFOLD
%%%%%%%%%%%%%%%%%%%%%%%%%%%%%%%%%%%%%%%%%%%%%%%%%%%%%
\begin{figure}
\refstepcounter{figure}
\label{figureConifold}
\begin{center}
\makebox[8cm]{
   \begin{turn}{0}%
     \epsfxsize=6cm
     \epsfysize=4cm
     \epsfbox{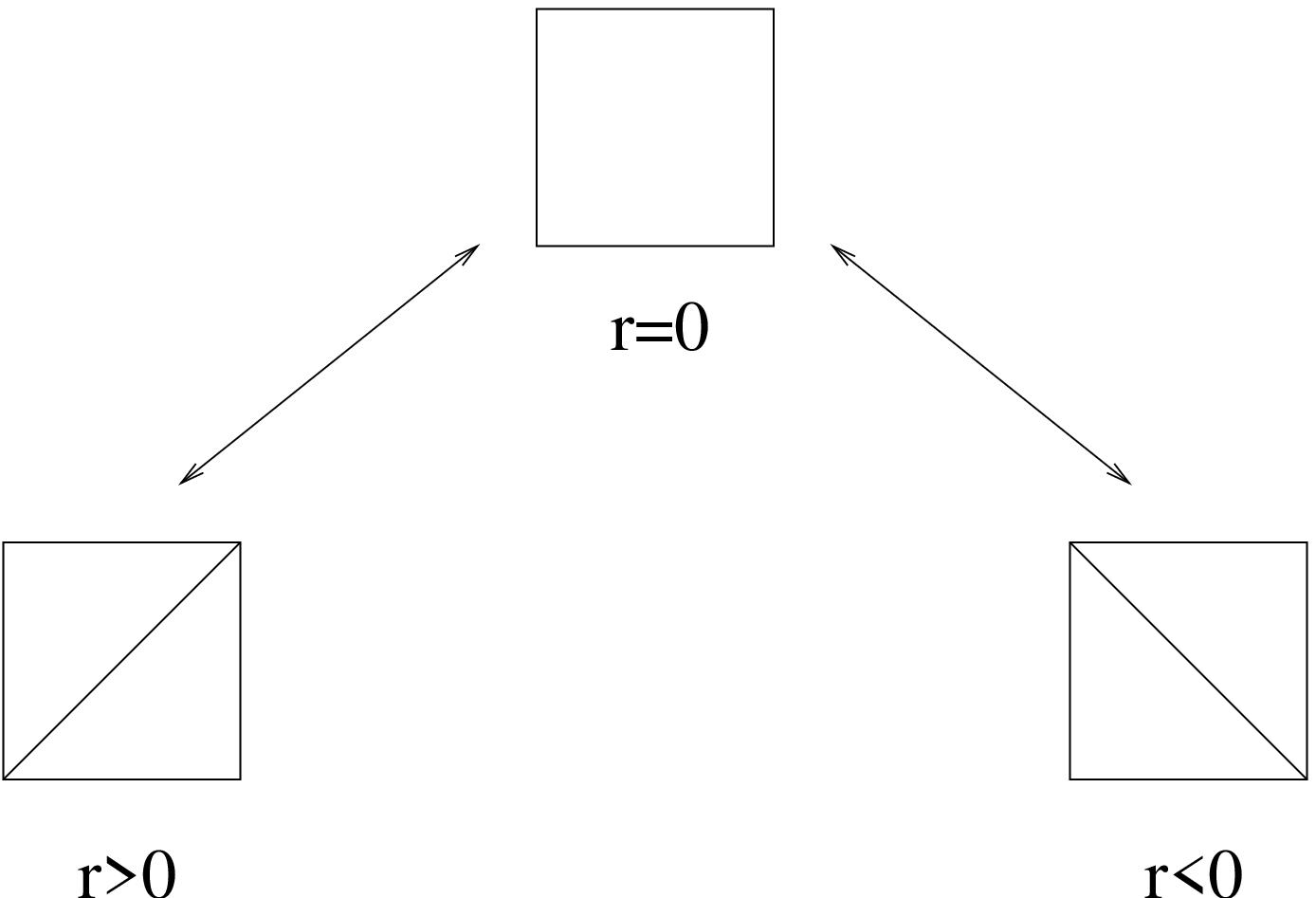}
   \end{turn}
}
\end{center}
\center{{\bf Fig.\thefigure}{\bf.}
             Two small resolutions of the conifold are related by a flop.}
\end{figure}
%%%%%%%%%%%%%%%%%%%%%%%%%%%%%%%%%%%%%%%%%%%%%%%%%%%%%
There clearly is a redundancy in this identification, since 
for any $\lambda \in {\mathbb C}^*$, taking $A_i \rightarrow \lambda A_i, B_j \rightarrow \lambda^{-1} B_j$
maps to the same point of the conifold. 
We can remedy this as follows. 
We will think about ${\mathbb C}^*$ as ${\mathbb{R}}_+\times S^1$,
that is we will put $\lambda = R e^{i \theta}$, with $R > 0$.
Take a quotient by ${\mathbb R}_+$ first, by picking $R$ to set
\begin{equation}
\label{con3}\quad  |A_1|^2 + |A_2|^2 - |B_1|^2 -|B_2|^2 = 0.
\end{equation}
To obtain a space isomorphic to the conifold we started with we must still 
divide by the $S^1 = U(1)$.\\
\noindent 
One can obtain a more physical interpretation of what we have done,
which stems from observation that the description of the conifold
we have come up with above is precisely that of a Higgs branch of a 
particular linear sigma model. It corresponds to a theory
with four real supercharges, 
gauge group $U(1)$ with four matter fields $A_i, B_j$
with charges $+1$ and $-1$, respectively and no superpotential. 
The D-flatness conditions are then given by equation (\ref{con3}).
This is of course 
not a new construction \cite{phases,klebwit}.\\
\noindent
Turning on the FI parameter $r$ will modify the D-flatness
conditions to 
\begin{equation}
\label{con4}\quad  |A_1|^2 + |A_2|^2 - |B_1|^2 -|B_2|^2 = r.
\end{equation}
We have three cases to consider here.
\begin{enumerate}
\renewcommand{\labelenumi}{\alph{enumi}.)}
\item  For $r=0$ we have a singular manifold 
       the conifold.
\item  For $r>0$, the origin $A_i=0=B_j$ of the conifold is replaced 
       by a sphere of size $ |A_1|^2 + |A_2|^2 = r$. From the point 
       of view of geometry, turning on the FI parameter \cite{phases} 
       is naturally interpreted as blowing up a sphere of size $r$.
\item  For $r<0$, from the point of view of b) the K\"ahler class is 
       negative. We do still have a smooth manifold, because now the 
       origin is replaced by $ |B_1|^2 + |B_2|^2 = r$.
\renewcommand{\labelenumi}{enumi}
\end{enumerate}
The manifolds in b.) and c.) are topologically distinct --
they are related by a flop transition (see Fig.\ref{figureConifold}).

\subsubsection*{Deformation}

In addition to the smoothings we discussed above, 
conifold singularity can be smoothed out by keeping the K\"ahler structure 
fixed but modifying the defining equation.
For this it suffices to change the complex structure to:
$$xy-uv = \epsilon.$$
As long as $\epsilon \neq 0$, the conifold singularity has been removed.
By examining the equation in detail, one can show that the origin was 
replaced by an $S^3$.

\subsection{More General Singularities}

We are now more or less in place to introduce toric geometry,
as a tool for treating more complicated singularities. 

We will use the language of linear sigma models to put the discussion on
a more physical basis \cite{phases,aspgree}. 
We are constructing a linear sigma model
whose moduli space will be a Calabi-Yau manifold ${\cal M}$.
First, the number of independent FI
parameters, or equivalently the number of $U(1)$ factors,
will equal $h^{1,1}({\cal M})$ (unless stated otherwise,
by ${\cal M}$ we mean the manifold obtained by smoothing out the
singularity). 
It is this number, and the charges of various matter multiplets that toric 
geometry must encode. 

A toric diagram consists of $d+n$ vectors $\{\vec{v}_i\}$ in a lattice 
${\mathbb N} = {\mathbb Z}^d$. 
Every vector $\vec{v}_i$ corresponds to a matter multiplet in our sigma 
model which we will call $x_i$.
To describe a toric variety homeomorphic to other than flat space
we need $n>0$. Since ${\mathbb N}$ is d-dimensional, there are $n$ relations between the 
$d+n$ vectors which we will write in the form
\begin{equation}
          \label{con9} 
          \quad \sum_{i=1}^{d+n} Q_i^a \vec{v}_i~=~0, 
          \quad a=1,{\ldots},n.
\end{equation}
It is clear that 
$Q$'s should be interpreted as the charges of the matter fields
under the $n$ $U(1)$'s. 
As a consequence the D-flatness conditions will read,
\begin{equation}
          \label{con5}
          \sum_{i=1}^{d+n} Q_i^a |x_i|^2~=~r_a, 
          \quad a=1,{\ldots},n.
\end{equation}
${\cal M}$ is a space of solutions to (\ref{con5}), up to the identifications
imposed by gauge symmetry. 
Or, instead of setting D-terms to zero and dividing by the 
gauge group, we could have taken a quotient by the complexified
gauge group\linebreak 
\hbox{$x_i \rightarrow \lambda ^{Q_i^a} x_i\,, \;\; a=1,{\ldots}, n\,$},
where $\lambda \in {\mathbb C}^*$
and express the moduli space as the space of gauge invariant polynomials in
$x's$, modulo any relations between them. This is the language of 
eq.(\ref{con}).

There is one slight simplification that occurs 
when ${\cal M}$ is a (non-compact) Calabi-Yau manifold.
Namely, ${\cal M}$ is a Calabi-Yau if and only if there exists
a vector ${\vec h} \in {\mathbb M}$, where ${\mathbb M}$ is the dual
lattice of ${\mathbb N}$, such that
\begin{eqnarray*}
        <\vec{h},\vec{v}_i>~=~1,\quad \forall\, \vec{v}_i
\end{eqnarray*}
i.e. if and only if all the vectors $\vec{v}_i$ live on a hyperplane 
a unit distance away from the origin of ${\mathbb N}$. Therefore in 
all of our examples toric singularities can be described by planar 
diagrams, only.

\subsubsection*{Hyperquotient Singularities}

As is well known, one
can obtain more complicated geometries by taking a quotient of the
simpler ones by a properly chosen group action. 
Dividing ${\mathbb C^n}$ by a discrete symmetry group
$\Gamma$ we obtain orbifolds with quotient singularities.
Taking a quotient of a hypersurface singularity like ${\cal C}$ we
obtain what are called hyperquotient singularities.
Both can be treated easily in the language of toric geometry.
First however, we must find appropriate symmetry group of our manifold.
Clearly, any action 
$x_i \rightarrow {\markfont \lambda_i} x_i, |\lambda| =1$ leaves
the manifold invariant. The symmetry group is $U(1)^{n+d}/U(1)^n =
\exp(2 \pi i \mathbb{Z}^d)$. 
More precisely, the toric variety ${\cal M}$
will contain the torus $T^d$ as a dense open subset.
There is a natural action of $U(1)^d = T^d$ on the toric variety
given as follows. To any element 
$\vec{n} \in {\mathbb{Z}}^{d}$, we can associate an element
of $U(1)^{d}$ via $x_i \rightarrow e^{i n_i \theta} x_i,$
where $\vec{n} = \sum n^i \vec{v}_i$ defined up to
$\sum Q_i^{a} \vec{v}_i=0$. 

So far, our lattice was integral. Now suppose we refine the lattice
by adding a vector in $\mathbb{Q}^d$ for example 
$\vec{q}=\frac{1}{r}(a_1,\ldots, a_d)$. For as long as the lattice
was integral the torus action was well defined. Now, it will be so only 
if we induce additional identifications on the $x_i$'s, namely
writing 
${\vec q} = \sum \frac{q_i}{r} {\vec{v}_i}, \pmod{\sum Q_i {\vec v}_i}$,
we should identify 
$$x_i \sim e^{\frac{2 \pi i q_i}{r}} x_i.$$ 
Perhaps a better way to express the action of the quotient,
is in terms of gauge invariant monomials.
Clearly, any ${\mathbb{C}}^*$ invariant monomial is of the form
$$x^m=\prod x_i^{<{\vec v}_i,{\vec m}>},$$
so the space of ${\mathbb{C}}^*$ invariant monomials is just the 
dual lattice ${\mathbb M}$. Actually we want a bit less, since
a) only the positive powers should appear, so we only want
those ${\vec m}$'s that satisfy $<\vec{m},{\vec v}_i>\geq 0$, 
$\forall i$, and 
b) we only want the independent ones.
Then, the identification induced on the monomials is
\begin{eqnarray*}
     x^m \sim e^{2 \pi i <{\vec q}, {\vec m}>} x^m.
\end{eqnarray*}
In any event, we should now be ready to produce
new spaces. We are up to producing orbifolds of the
conifold, ${\cal C}/{\Gamma}$. Let us take $\Gamma = {\mathbb Z}_k
\times {\mathbb Z}_l$. 
So, start with our conifold ${\cal C}$, defined by four vectors
$\vec{v}_{1,2,3,4} \in {\mathbb N}$ as before, but refine the 
lattice to ${\mathbb N}'$ by adding two vectors, 
${\vec e}_{k} = (\frac{1}{k},0,0)$, and
\hbox{${\vec e}_ {l}= (0,\frac{1}{l},0)$}.
The resulting toric diagram (cf. Fig.\ref{figureRectanglekxl}) 
``looks'' the same
as that for the conifold ${\cal C}$, except for the fact that it lives
in a finer lattice.  
This, as explained above, results in the following identifications:
$${\vec e}_k=\frac{1}{k}({\vec v}_2-{\vec v}_1),$$
we find that the quotient acts by
$$ A_1 \sim e^{-\frac{2 \pi i}{k}}  A_1, \quad 
B_1\sim  e^{\frac{2 \pi i}{k}}B_1,\quad
A_2 \sim A_2, \quad 
B_2\sim  B_2,$$ 
and similarly for ${\vec e}_l$,
$${\vec e}_l=\frac{1}{l}({\vec v}_3-{\vec v}_1),$$
the quotient is by
$$ A_1 \sim e^{-\frac{2 \pi i}{l}}  A_1, \quad 
B_1\sim  B_1,\quad
A_2 \sim A_2, \quad 
B_2\sim e^{\frac{2 \pi i}{l}} B_2.$$ 
Equivalently on $xy = uv$, we identify 
$x \sim x, y \sim y, u \sim e^{-\frac{2 \pi i}{k}} u,
v \sim e^{\frac{2 \pi i}{k}} v$, and $x \sim  e^{-\frac{2 \pi i}{l}}x, 
y \sim e^{\frac{2 \pi i}{l}} y, u \sim  u,
v \sim  v.$
In terms of $\Gamma$ invariant coordinates
$x' = x^l, y' = y^l, z = x y, u'=u^k, v'=v^k, w = uv,$
the defining equation of the conifold becomes simply $z=w$. 
Taking into account
that not all the invariant monomials are independent, the 
$\Gamma = \mathbb{Z}_k \times \mathbb{Z}_l$ orbifolded conifold, after obvious
renaming of variables 
becomes:
\begin{equation}
\label{zkxkp} {\cal C}_{k,l}:\quad xy=z^l, \quad uv=z^k. 
\end{equation}

%%%%%%%%%%%%%%%%%%%%%%%%%%%%%%%%%%%%%%%%%%%%%%%%%%%%%
%   BILD: RECTANGLE WITH K X L
%%%%%%%%%%%%%%%%%%%%%%%%%%%%%%%%%%%%%%%%%%%%%%%%%%%%%
\parbox{12cm}{
  \refstepcounter{figure}
  \label{figureRectanglekxl}
  \begin{center}
  \makebox[6cm]{
   \begin{turn}{0}%
     \epsfxsize=6cm
     \epsfysize=4cm
     \epsfbox{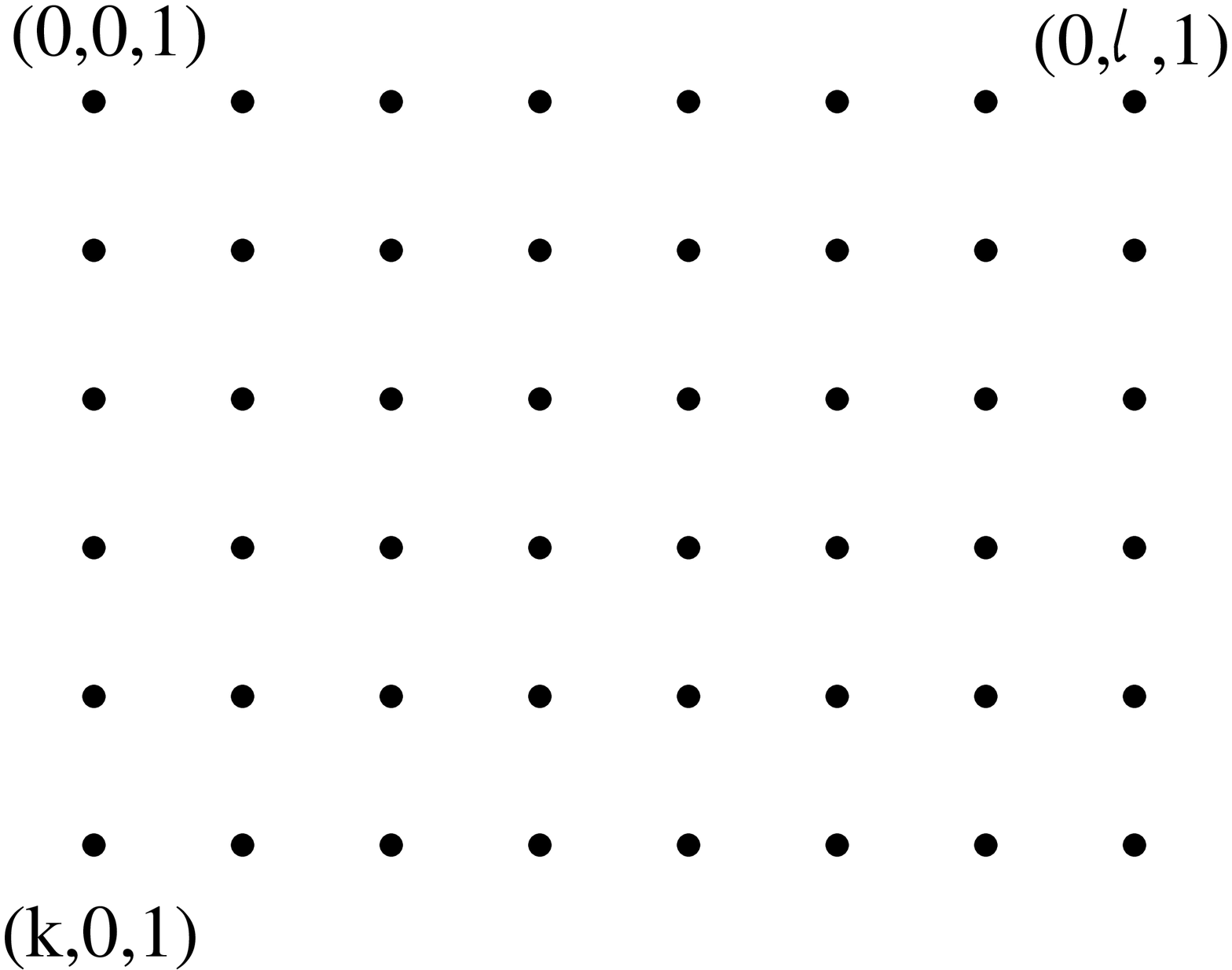}
   \end{turn}
  }
  \end{center}
  \center{{\bf Fig.\thefigure}{\bf.}
             (Blowup of) orbifolded conifold ${\cal C}_{k,l}$.}
}
%%%%%%%%%%%%%%%%%%%%%%%%%%%%%%%%%%%%%%%%%%%%%%%%%%%%%

\subsubsection*{Blowing Up}
Toric geometry has equipped us with a means of blowing up the singularity.
First let's look at the orbifold ${\cal C}_{kl}$. There are still only four vectors
defining the diagram which were inherited from the conifold.
There is a single relation between them, and thus a single K\"ahler class
but this is insufficient to smooth out
${\cal C}_{k,l}$. However, due to the fact that
the lattice is finer, there exist lattice points within the rectangle,
these are. all the points 
${\vec v}_{i,j} = (i,j,1)$, $0\leq i \leq k$, $0\leq j\leq l$.
We can add these points to the toric diagram. In the language of linear sigma models, the effect is to add more matter 
fields, but also more $U(1)$ factors, and thus more FI parameters. 
Clearly, the resolved manifold will have
$h^{1,1}({\cal C}_{k,l}) = (k+1)(l+1)-3$,
which is the total number of linearly dependent vectors within the diagram.
(Points outside the diagram can be added as well. 
However they will not contribute to the resolution
of the singularity, but only modify it by irrelevant pieces.)
We will not try to specify precise region in the 
K\"ahler structure moduli space where the resolution lives, which would
correspond to picking a triangulation of the toric diagram,
because we will not need this piece of information.
It is clear there will be very many 
different such regions, and they are all related by flops.

Finally, starting from the orbifolded conifold ${\cal C}_{k,l}$, with $k$, $l$ 
sufficiently large, by performing partial resolutions
we can obtain essentially any other toric 
singularity\footnote{These singularities have been
introduced in the physics literature for the description of gauge theories
in \cite{klemm1}.}.
The
basic fact to note that adding or subtracting one of the boundary
points of the diagram changes $h^{1,1}\rightarrow h^{1,1}-1$.
The right interpretation of this is that we are probing the
region of the K\"ahler structure moduli space 
where the four cycle associated to this point in the toric diagram
becomes
large enough, that it in fact becomes irrelevant to the local physics --
the associated vectors can be dropped altogether.

We will provide some more examples of the spaces we
will explicitly use in this paper and introduce some terminology.

Starting from an orbifold of ${\cal C}$ by ${\mathbb Z}_k$,
\begin{equation}
\label{zk} {\cal C}_k:\quad xy=z^k, \quad u v = z,
\end{equation}
or equivalently $xy=(uv)^k$, which has $h^{1,1}({\cal C}_k)=  2k-1$,
by partial resolution we can obtain
the generalization of a conifold,
\begin{equation}
\label{coni} {\cal G}_{kl}:\quad x y=u^k v^l
\end{equation}
with only 
$k+l-1$ K\"ahler structure deformations.
Clearly, in our notation ${\cal C}_k\equiv {\cal G}_{kk}$.\\
\parbox{12.5cm}{\hspace{1cm}\parbox{6.0cm}{
    %%%%%%%%%%%%%%%%%%%%%%%%%%%%%%%%%%%%%%%%%%%%%%%%%%%%%
    %   BILD: RECTANGLE WITH K + l
    %%%%%%%%%%%%%%%%%%%%%%%%%%%%%%%%%%%%%%%%%%%%%%%%%%%%%
    \refstepcounter{figure}
    \label{figureRectanglek+l}
    \begin{center}
    \makebox[5cm]{
      \begin{turn}{0}%
        \epsfxsize=4cm
        \epsfysize=1.5cm
        \epsfbox{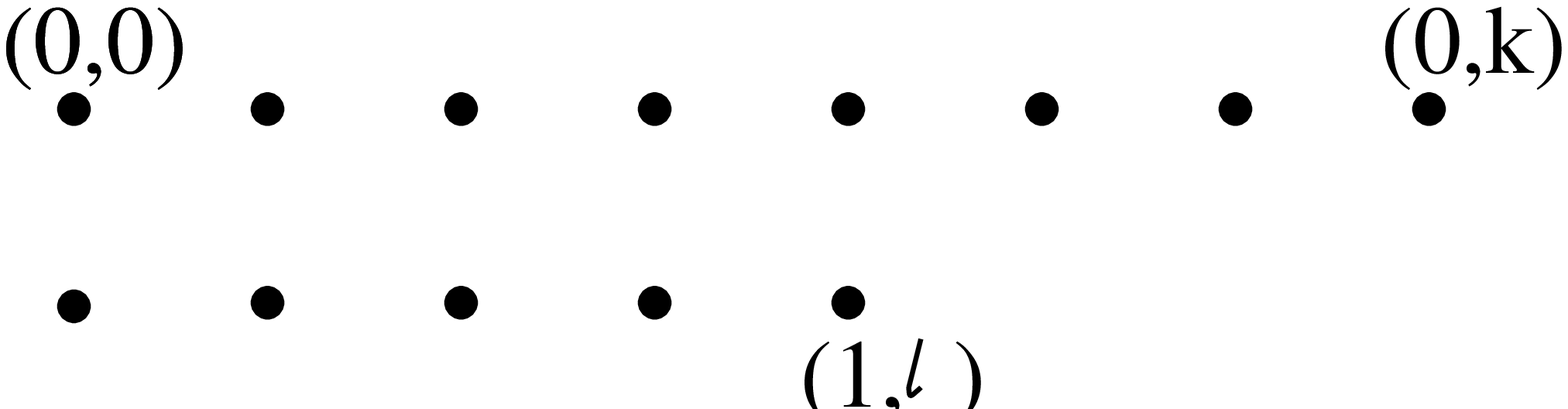}
      \end{turn}
    }
    \end{center}
    \vspace{3.5ex}
    \center{{\bf Fig.\thefigure}{\bf.}
             (Blowup of) generalized \\ conifold ${\cal G}_{kl}$.}
    %%%%%%%%%%%%%%%%%%%%%%%%%%%%%%%%%%%%%%%%%%%%%%%%%%%%%
    %
    %%%%%%%%%%%%%%%%%%%%%%%%%%%%%%%%%%%%%%%%%%%%%%%%%%%%%
}\hfill\parbox{5.0cm}{
    %%%%%%%%%%%%%%%%%%%%%%%%%%%%%%%%%%%%%%%%%%%%%%%%%%%%%
    %   BILD: TRIANGLE
    %%%%%%%%%%%%%%%%%%%%%%%%%%%%%%%%%%%%%%%%%%%%%%%%%%%%%
    \refstepcounter{figure}
    \label{figureTriangle}
    \begin{center}
    \makebox[5cm]{
        \begin{turn}{0}%
          \epsfxsize=4cm
          \epsfysize=3cm
          \epsfbox{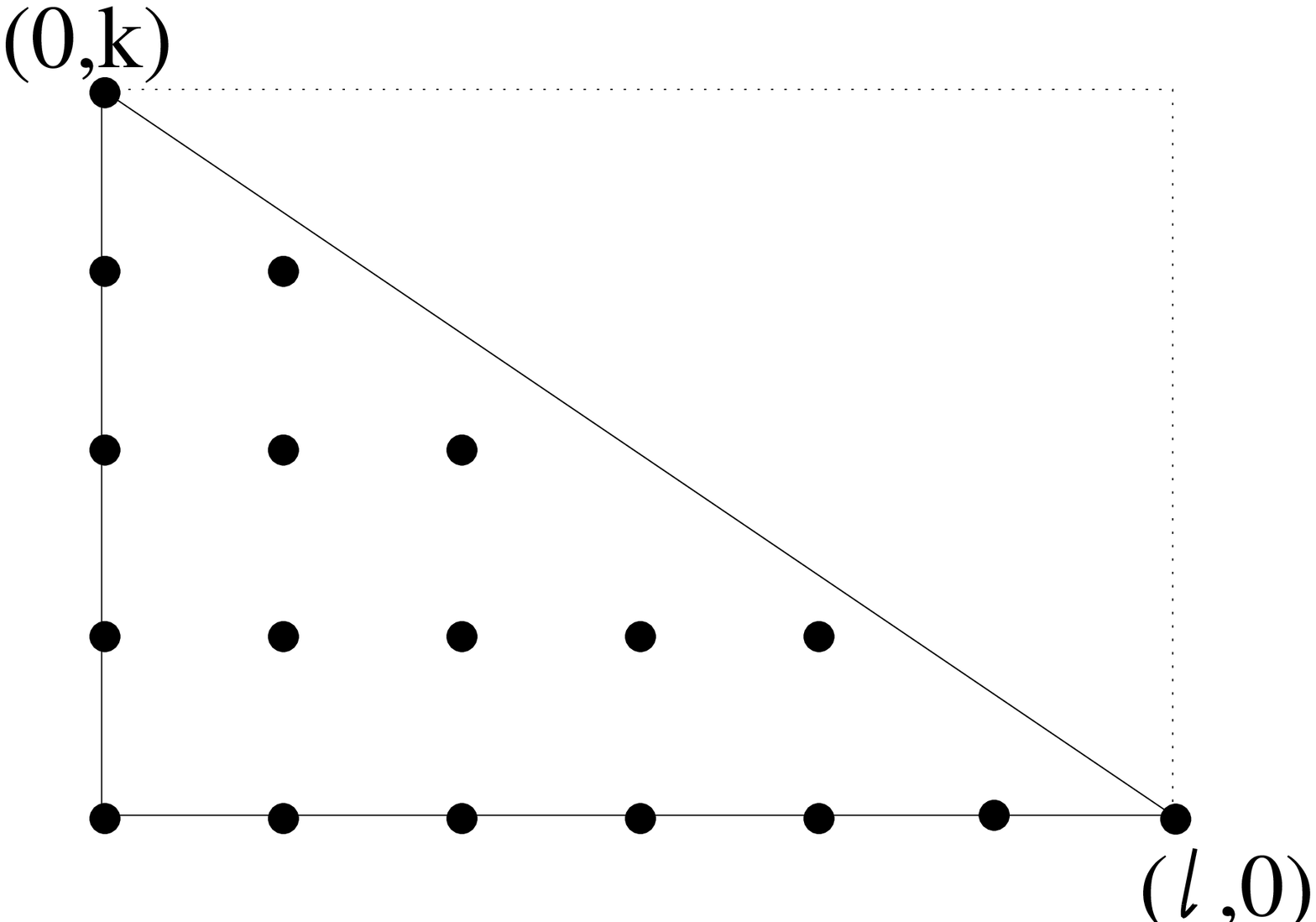}
        \end{turn}
    }
    \end{center}
    \center{{\bf Fig.\thefigure}{\bf.}
             Toric diagram of\\ the
             ${\mathbb{Z}}_k\times {\mathbb{Z}}_l$-orbifold ${\cal O}_{kl}$.}
    %%%%%%%%%%%%%%%%%%%%%%%%%%%%%%%%%%%%%%%%%%%%%%%%%%%%%
    %
    %%%%%%%%%%%%%%%%%%%%%%%%%%%%%%%%%%%%%%%%%%%%%%%%%%%%%
}\vspace{0.4cm}}\\
The conventional orbifold ${\cal O}_{kl}=
{\mathbb C}^3/{\mathbb Z}_k\times{\mathbb Z}_l$
can be found in the K\"ahler structure moduli space of the orbifolded conifold
${\cal C}_{kl}$, the toric diagram of the orbifold being
contained in that of the orbifolded conifold.
One way to see this is indeed a toric diagram corresponding
to ${\mathbb C}^3/{\mathbb Z}_k\times{\mathbb Z}_l$ is to use the fact
that the diagram can be obtained starting from a toric diagram containing
just three vectors $\vec{v}_1=(0,0,1),\vec{v}_2=(1,0,1),\vec{v}_3=(0,1,1)$
in an integral lattice ${\mathbb N}$, which gives a toric variety 
homeomorphic to flat space, and then refine the lattice to
${\mathbb N'}$, as in  Fig.\ref{figureTriangle}. 
The map from the toric variety 
in ${\mathbb N}$ to the one living in ${\mathbb N}'$ is one to one provided
one includes discrete identifications on
the three matter fields $A_i, i=1,2,3$,

$$A_1\sim e^{-\frac{2\pi i}{l}}A_1, 
\quad A_2\sim e^{\frac{2\pi i}{l}}A_2, \quad A_3\sim A_3,$$
and
$$A_1\sim e^{-\frac{2\pi i}{k}}A_1, \quad A_2\sim A_2, \quad A_3\sim
e^{\frac{2\pi i}{k}}A_3.$$

As before,
number of K\"ahler 
structure deformations is just the number of independent points in the toric 
diagram, and this number will clearly depend on whether
$(k,l)$ are coprime or not, since the number of points on the diagonal is
$gcd(k,l)+1$. 

\subsubsection*{Deformations}
\begin{itemize}
\item The orbifolded conifold  ${\cal C}_{kl}: \quad xy=z^k,
\quad uv=z^l$ can be deformed to a smooth space by modifying the defining 
equation as:
\begin{eqnarray}
   x y~=~\prod_{i=1}^k (z-w_i)\quad\quad u v~=~\prod_{j=1}^l 
   (z-w'_j).\label{zkxkpd}
\end{eqnarray}
One of these parameters can be set to 1 by shifting $z$,
so we are left with $k+l-1$ parameters.
This gives $h^{2,1}({\cal C}_{kl})=k+l-1$.\\
%\noindent
%The blowups of the orbifolded conifold are encoded in the
%toric diagram in Fig.\ref{figureRectanglekxl}. We see $(k+1)(l+1)$
%K\"ahler parameters $m_{ij}$ (3 of which can be set to 1).\\
\end{itemize}

\begin{itemize}
\item The generalized conifold ${\cal G}_{kl}: 
\quad xy=u^k v^l$ can be deformed into
\begin{equation}
\label{conid} xy= \sum_{i,j=0}^{k,l} m_{ij} u^i v^j.
\end{equation}

This time we see $h^{2,1}({\cal G}_{kl})=(k+1)(l+1)-3$ 
complex structure deformations $m_{ij}$:
by shifting $u,v$, we can eliminate two of the parameters, and another one
by rescaling the defining equation.
\end{itemize}

\subsubsection*{Mirror Symmetry}

Toric geometry is well adopted to discussing mirror symmetry as well.
We will review it here very briefly, only.
Mirror symmetry exchanges the K\"ahler structure parameters with the
complex structure parameters.
Now, to understand the mirror map, we first need to 
know something about the complex 
structure moduli space.
How is the complex structure encoded in the equation of the manifold?
The answer is as follows:  the coefficients of the monomials
appearing in the defining equation are coordinates on the complex
structure moduli space. What they parameterize are the ``sizes''
of various three-cycles (i.e. the periods of the holomorphic three form)
and the metric on the moduli space.
The periods, (and therefore the metric -- the moduli space has special 
geometry structure)
can be derived directly as a solution
to a system of differential equations. The main point is that
the differential equations depend solely on the relationships
between the monomials in the defining equation and nothing else.

Given the toric manifold, relations between the vectors in the toric diagram
of ${\cal M}$
(we are assuming a completely smooth space here, with all 
the possible blowups performed),
$$\sum_{i=1}^{n+d} {Q_i}^a {\vec v}_i, \quad a=1,\ldots n$$
map to relationships between the monomials in the defining equation of
${\cal W}$, the mirror of ${\cal M}$, given by
\begin{equation}
\label{defining}
{\cal W}: \quad     \sum_i a_i m_i=0,
\end{equation}
where $a_i$ are coefficients, and $m_i$ monomials, the monomials must satisfy
\begin{equation}
\label{monomial}
\prod_{i=1}^{n+d} m_i^{Q_i^a}=1, \quad a=1,\ldots n.
\end{equation}

Any solution to these equations (and in general there are more than one)
will represent the same complex structure (by decoupling of complex and
K\"ahler moduli spaces).
Note that there are $n+d$ monomials with $n$ relations between them.
Together with the hypersurface equation, this gives a $d-1$
dimensional manifold, but the homogeneity of the monomial relations will allow
us to remove one more. The mirror will naively have $d-2$
dimensions. This 
is not a problem, rather an artifact of the fact 
that local mirror symmetry is
encoding all the information about the complex structure of the mirror, and nothing but. One can fix the ``dimensionality'' of the local
mirror by  adding quadratic pieces, as this will not influence the
complex structure moduli space.

Let us briefly show how this works for the two examples we are going
to be concerned with in this work
\footnote{These examples and many more along these lines have been recently 
analyzed in great detail in \cite{klemm2}.}, 
${\cal G}_{kl}$ and ${\cal C}_{kl}$.
Consider first the blowup of ${\cal C}_{kl}$. We want to interpret
the same diagram Fig.\ref{figureRectanglekxl} as defining the complex
structure of the mirror. Assigning the vector $(i,j,1)$ to a monomial
$u^i v^j$ clearly eq.(\ref{monomial}) is satisfied for all the
relations. The defining equation for the mirror of ${\cal C}_{kl}$
hence becomes according to eq.(\ref{defining}):
$$ \sum_{i,j=0}^{k,l} m_{ij} u^i v^j =0$$
or after adding the irrelevant quadratic piece $xy$
$$ xy= \sum m_{ij} u^i v^j $$
which is nothing but the deformation of ${\cal G}_{kl}$.

Having established that the deformation of ${\cal G}_{kl}$ is mirror
to the blowup of ${\cal C}_{kl}$ we can find another dual pair
by following our geometries through a conifold transition.
We should find that the blowup of
${\cal G}_{kl}$ is the mirror of the deformation of ${\cal C}_{kl}$.
Let us see how this works. As above we read of the mirror to be
$$ \prod_{i=1}^k (z-w_i) + t \prod_{j=1}^l (z-w_j') =0.$$
Because $t$ appears only linearly
this encodes the same complex structure as
$$ \prod_{i=1}^k (z-w_i) = uv, \quad \prod_{j=1}^l (z-w_j')=xy $$
which is indeed the deformation of ${\cal C}_{kl}$ as
presented in eq.(\ref{zkxkpd}).

%%%%%%%%%%%%%%%%%%%%%%%%%%%%%%%%%%%%%%%%%%%%%%%%%%%%%%%%%%%%%
%       SECTION:  GAUGE THEORIES
%%%%%%%%%%%%%%%%%%%%%%%%%%%%%%%%%%%%%%%%%%%%%%%%%%%%%%%%%%%%%

\section{The gauge theories}
\label{sectiongauge}

Having introduced the geometric background spaces, we will now
discuss the corresponding gauge theories if one adds $M$ D3 branes with
world volume transverse to the non-compact manifolds. 
The corresponding gauge group for the orbifolded conifold 
${\cal C}_{kl}$, eq.(\ref{zkxkp}),
is given by the following $N=1$ supersymmetric gauge theory:
\begin{eqnarray} \label{conigauge}
SU(M)^{kl} &\times& SU(M)^{kl}
\end{eqnarray}
with matter fields
$(A_1)_{i+1,j+1;I,J}$, $(A_2)_{i,j;I,J}$, $(B_1)_{i,j;I,J+1}$,
$(B_2)_{i,j;I+1,J+1}$. We label the gauge groups with $i,I=1 \ldots
k$ and $j,J=1 \ldots l$. All the matter fields are bifundamental under
the gauge groups indicated by the indices. The just $\mathbb{Z}_k$ orbifolded
conifold arises as the special case $l=1$.
In addition there will be a quartic superpotential
\begin{eqnarray}
\label{sup4}
W&=&\sum_{i,j} (A_1)_{i+1,j+1;I,J} (B_1)_{i,j;I,J+1}
(A_2)_{i,j+1;I,J+1} (B_2)_{i,j+1;I+1,J+1}\nonumber\\ &-&
\sum_{i,j} (A_1)_{i+1,j+1;I,J} (B_1)_{i,j;I+1,J}
(A_2)_{i+1,j;I+1,J} (B_2)_{i+1,j;I+1,J+1}.
\end{eqnarray}
The other singularity, the generalized conifold eq.(\ref{coni}) corresponds to
$$SU(M)^{k+l}$$
with bifundamental matter according to Uranga's rules \cite{Uranga} 
and quartic 
superpotentials.

Finally consider $M$ D3 branes on a transversal orbifold singularity
${\cal O}_{kl}$. 
They give rise to an
\begin{equation}
\label{boxgauge}
SU(M)^{kl}
\end{equation}
gauge theory
with 3 types of chiral bifundamental multiplets $H_{i,j;i+1,j}$, \linebreak
$V_{i,j;i,j+1}$
and $D_{i+1,j+1;i,j}$ in each gauge group and
a cubic superpotential
\begin{equation}
\label{supo3}
W=\sum_{i,j} H_{i,j;i+1,j} V_{i+1,j;i+1,j+1} D_{i+1,j+1,i.j}- 
\end{equation}
$$ \sum_{i,j} V_{i,j;i,j+1} H_{i,j+1;i+1,j+1} D_{i+1,j+1;i,j}.$$

This way
the orbifold gauge theories will
have $3M$ matter fields per gauge group and cubic superpotentials, leaving
us with a finite theory. The conifold gauge theories have $2M$ matter
fields per gauge group and quartic superpotentials. These theories are
non-finite but flow to a fixed line parameterized by a marginal operator in the
IR.

%%%%%%%%%%%%%%%%%%%%%%%%%%%%%%%%%%%%%%%%%%%%%%%%%%%%%%%%%%%%%
%       SECTION:  MIRROR SYMMETRY
%%%%%%%%%%%%%%%%%%%%%%%%%%%%%%%%%%%%%%%%%%%%%%%%%%%%%%%%%%%%%

\section{The T-dual brane setups and mirror symmetry}
\label{sectionmirror}

In this section we would like to discuss the brane configurations which
are T-dual to the singularities introduced in 
section \ref{sectiongeometries}.
Specifically, we are interested in two different T-duality
transformations: the first one was recently
discussed by Uranga \cite{Uranga} and by Dasgupta and Mukhi 
\cite{dasguptamukhi}.
The dual brane picture consists of NS
and rotated
NS' 5-branes we will henceforth refer to it as $T_U$. 
The D3 branes which we want to study in the next section
become D4 branes after the $T_6$
duality transformation which live on the compact interval in $x^6$.

Second we perform a T-duality along the compact directions $x^4$ and $x^8$,
$T_{48}=T_4T_8$. This maps the singularities again to NS and NS' branes,
where now the D3 probes become D5 branes which fill the compact brane
box in the $x^4-x^8$ spatial directions. This T-duality was first introduced
in \cite{DonD} and for a special point in moduli space used by
\cite{haur} to study D3 branes on orbifold singularities. We will henceforth
refer to it as $T_V$.

These T-dualities are very useful in the sense that they allow us to read
of the gauge groups on the D3 brane world volume according to some very
intuitive graphic rules encoded in the brane configuration. While
for the orbifold a perturbative string calculation is also available to
get the gauge group, for the more general singularities discussed here,
one would have to rely on a technique as in
\cite{klebwit}.

Combining the two, that is doing $T_{468}$ we actually performed
a local mirror symmetry transformation.
We will see explicitly, that $T_{mirror}$ takes a geometry ${\cal W}$
 into its mirror 
geometry ${\cal M}$. 
The gauge theory of a D3 brane probing ${\cal W}$ has to be identical 
to that on  a D6 brane wrapping a 3-cycle in ${\cal M}$.
  
This should correspond to the mirror transformations
for Calabi-Yau spaces, which are the compact counterparts of
our non-compact singularities. These compact CY's are assumed 
$T^3$-fibrations (with $T^3$ a special Lagrangian submanifold of the CY) 
and the mirror transformations acts as the inversion of the volume of the
$T^3$. Obviously, this $T^3$ corresponds to our three 
directions $x^4$, $x^6$ and $x^8$, on which the mirror symmetry acts.

\subsection{The brane setup}

Before we embark on the discussion let us briefly recall the basic
brane setup. There are two configurations we are going to
consider, for one the standard HW \cite{HW,EGK} type of brane setup,
where D4 branes are stretched in between NS and NS' branes, former
living along 012345 and latter along 012389, the rotation
being necessary in order to break SUSY down from 8 to 4 supercharges.
In order to have a supersymmetric theory from D4 branes on the interval
all the NS and NS' branes have to be at the same position in the 7 direction.
Separations along the 7 direction would be interpreted as FI terms or
baryonic branches in the gauge theory and effectively leads to a breaking
of the gauge group we want to see.
Similarly we should require all the NS branes to have the same position
in 89 and all the NS' branes to have the same position in 45 space.
They are separated along the 6 direction building the intervals, 
along which the D4 branes (living in 01236) stretch.

The second kind of brane setup we are going to consider are the so called brane
boxes \cite{hanzaf}, which are a straight forward generalization of
the interval theories. 
The brane box is a rectangle bounded by NS and NS' branes with a D5 brane
suspended on it. This can be achieved
by the same NS and NS' branes as above but now all branes have to
be located at the same 67 position, closing the intervals. We can open
up the boxes by separating the NS and NS' branes along their 48 directions
(unfortunately this way we differ from the notation in \cite{hanzaf},
where the boxes were taken to live in the 46 space. This is necessary,
since it is crucial for us, that box and interval can be realized by the
same set of NS and NS' branes). We still want to keep the 5 and 9 positions
equal in order to preserve supersymmetry of the suspended probes. 
Deformations along these
directions are again FI terms in the gauge theory, which are reinterpreted
as baryonic branches after freezing out the diagonal $U(1)$s.

\subsection{Deformations and blowups}

As mentioned above, it is important to distinguish whether we want to study
the deformation or the blowup of the singularity under investigation.
The corresponding parameters should have an interpretation in the brane 
picture as well. 
If the dual is `pure brane', i.e. consists only of branes in 
 flat space, this interpretation will be solely in terms of NS brane positions 
 and, as will be established later, on brane shapes.  
Otherwise some of the parameters encode blowups of the non-trivial 
background geometry.
Even though latter description of the probe may still be useful, 
e.g. in order to read off gauge group and matter content, we would like to 
focus in the rest of our discussion on the case, where the dual is 
`pure brane'.

Let us forget for a moment about the D brane probes altogether. That is, we
want to study the map of the singular geometry into a configuration of NS
branes, as pioneered in
\cite{DonD}. Actually it turns out to be easier to start with
the NS brane configurations, where it is clear what we mean by
the 4, 6 and 8 direction. Performing $T_{48}$ and $T_6$
respectively we will find two different geometries, which 
have to be the local mirrors of each other. By construction
these are precisely the geometries that have a pure brane dual (we started
out with a pure brane setup!). We will find the following relations, as 
indicated in Fig.\ref{figureSummary} in the summary at the end of
this paper:

\begin{itemize}
\item The blowup of the generalized conifold is $T_U$ dual to
NS branes separated along 67 (the interval). These are in turn
$T_V$ dual to the mirror, the deformation of the orbifolded conifold.
\item Similarly the blowup of the orbifolded conifold will
$T_V$ dualize into a box and then $T_U$ dualize in the mirror,
the deformation of the generalized conifold. 
\end{itemize}
Indeed these two
transformations are related by a conifold transition, that is
bringing together the NS branes on the interval and then separating them 
along 4589 instead corresponds to blowing down the 2-cycles and opening up
the 3-cycles of the deformed conifold (and vice versa for the orbifolded
conifold).

\subsection{The brane box, blowup of the orbifolded and deformation
of the generalized conifold}

Let $m_i = (x^8,x^9)$, $m'_j=(x^4,x^5)$ positions of the $k$ NS and $l$
NS' branes respectively in $x^{4,5,8,9}$, and $w_i=(x^6,x^7)$, $w_j'=(x^6,x^7)$ 
the positions in the other two directions.

\noindent
Let us start with a ``brane box'', that is
we set all the $w_i$ and $w_j'$ to zero.
T-dualizing the brane box along $x^{4,8}$ we obtain a manifold we
call ${\cal M}$ and T-dualizing along $x^6$ we obtain ${\cal W}$.
The resulting geometries are related by 
$T_{468}=T_{mirror}$. 

\begin{itemize}
\item ${\bf T_V=T_{48}:}$  
           The T-dual space ${\cal M}$ is a $\mathbb{Z}_k\times\mathbb{Z}_l$ 
           orbifolded conifold 
           $${\cal C}_{k,l}: \quad xy=z^l, \quad uv=z^k$$ as in (\ref{zkxkp}), 
           where $k$, $l$ are numbers of NS and NS' branes.
	   This is a double ${\mathbb{C}}^*$ fibration over the $z$ plane,
	   that is the space has 2 U(1) isometries used in T duality.
	    The $x^4$, $x^8$ separations of the branes
	   must map into B-fluxes through 2-cycles of the T-dual space.
           We must therefore identify $m_i$, $m_j'$ as deformations of the 
	   {\sl K\"ahler} structure.
Deformations of the K\"ahler structure cannot change the complex structure,
so the $m_i$ and $m_j'$ will not be visible in the defining equations.
Having identified $m_i$, $m_j'$ as the K\"ahler structure parameters,
$w_i$ and $w_j'$ are identified as complex structure parameters.
But they are frozen, since turning them on would destroy the box structure.

For definiteness
take IIB theory on ${\cal C}_{kl}$.
T$_V$ duality takes us back to type IIB with NS branes.
In this case K\"ahler structure parameters, that is the 2-sphere sizes,
 sit in hypermultiplets. The other 
3 scalars in this multiplet are the NS-NS B-flux the RR B-flux and the 
RR 4-form-flux through the sphere.
Latter is a 2-form in 4d, which can be dualized into a scalar.
The 2-sphere size and the NS-NS B-flux are the complexified
K\"ahler parameter which map to $m_i$ and $m_j'$ under T$_V$.
In the brane box the two other scalars 
come from
Wilson lines of the NS-world volume gauge fields in 45 and 89 which
pair up in hypermultiplets with $m_i$ and $m_j'$ respectively.

Note however that we have a puzzle. 
The orbifolded
conifold ${\cal C}_{kl}$ has, as we have found from the toric
description,
$$(k+1)(l+1)-3=kl +k+l-2$$ K\"ahler structure parameters $m_{ij}$
which can be turned on to smooth out ${\cal C}_{kl}$.
 Only $k+l-2$ have been
realized in terms of the (relative) brane positions $m_i$ and $m_j'$.

So where are the $kl$ hypermultiplets in the brane box skeleton?
They sit at the $kl$ intersections! Strings stretching from NS to NS'
give rise to precisely these hypermultiplets 
\footnote{They are Strominger's D3 brane
on the vanishing 3-sphere in the geometry (remember that we only consider
blowups, so the 3-spheres are fixed at zero size).}. 

Turning on vevs for the two scalars corresponding to 2-sphere
sizes and NS-NS fluxes
resolves the intersection of the NS and NS' into a smooth
object, a little `diamond'. For
non-zero 
B-fields this diamond will open up in the 48 plane, for 2-sphere sizes in
the 59 plane. This interpretation will become more
suggestive after
discussing $T_U$ on this configuration and once we start discussing
the D3 brane probes. 

In the geometry the 2-spheres give rise to strings from wrapping D3 branes
around them. How do we see them in the NS5 box skeleton? The D3
branes on the $k+l-2$ spheres from
the curves of singularities correspond to (fractional)
D3 branes living in the
boxes (or better in whole stripes). The additional $kl$ strings must
now correspond to D3 branes {\sl in the diamonds}. We will indeed see
that the diamonds allow for such a configuration.

Of course the same story can be repeated in type IIA. Here the diamonds
will correspond to matter on the intersection of type IIA NS5 branes,
this time sitting in a vector multiplet.
Again the 2 scalars correspond to the $kl$ sizes and B-fluxes of the 
corresponding 2-spheres. Instead of the two additional
scalars in the hypermultiplet we this time see a vector from
the RR 3-form on the sphere. In the brane language the Wilson lines
of the NS5 gauge field
have to be substituted by Wilson lines of the (2,0) 2-form field, again
giving rise to vectors.

\item ${\bf T_U=T_6}$, T-duality to ${\cal W}$. What happens now is as follows.
      Since we did a $T_6$ duality, $x^6$ separations will
      become the B-fields. Thus, now the $w_i$ (which
      had to be put to zero since we are discussing a box) parameters are
      {\sl K\"ahler} structure deformations, while the non-zero
      $m_{ij}$ now should show up as complex structure deformations.

      The dual geometry should be a single ${\mathbb{C}}^*$ fibration.
      This will be described by an equation whose parameters,
      the complex structure deformations, must be $m_{ij}$.
      Let us first study the situation where the vevs of the
hypers living at the intersections are zero. In this case
      the ${\mathbb{C}}^*$ fibration must degenerate over
the NS and NS' positions
      $m_i$, $m_j'$, but in an independent way, since the branes are
      orthogonal -- it must contain two curves
      of singularities $A_{m-1}$, and $A_{n-1}$ corresponding to NS and NS'
      branes. There is one such equation for generic values of
      $m_i$'s
      $$ {\cal W}:
      \hspace{0.3cm}  uv= \prod_{i=1}^k (z-m_i) \prod_{j=1}^l (w-m_j')$$
The curve contains $kl$ conifold singularities located at
$z=m_i$ and $w=m_j'$ corresponding to the fact that all the
hypermultiplets at the intersections where turned off.

      Let us jump ahead and let us realize ${\cal W}$ directly
      as the mirror of ${\cal M}$. Performing the local mirror map
      we obtain:
      
      $$ {\cal W}:
      \hspace{0.3cm} uv=\sum_{i=0}^k \sum_{j=0}^{l} m_{ij} z^i w^j.$$
      By now the T-dual interpretation of this more general space should 
      be clear. It describes a single NS brane wrapping a curve
   
      $$\Sigma:\hspace{0.3cm}0=\sum_{i=0}^k \sum_{j=0}^{l} m_{ij} z^i w^j.$$
      The smoothing out of the intersections corresponds to the diamonds.
      For example one intersecting NS and NS' brane is described
      by $zw=0$. Turning on the hypermultiplet corresponds
      to smoothing this out to $zw=m_{00}$, as e.g discussed in \cite{sengp} 
for the related case of intersecting D7 branes.
      Indeed 
      the resulting smooth curve has a non vanishing circle of radius 
      $(m_{00})^{1/2}$ as can be seen by writing it as $x^2+y^2=m_{00}$ 
      and restrict oneself to the real section thereof, 
      for example\footnote{
           We are very grateful to M. Bershadsky for very
           helpful discussions on this point}. 
      This is precisely what we need: we can suspend a D3 brane as a
      soap 
      bubble on the NS skeleton, its boundary being given by the circle. 
      The tension of the resulting string is given by the area of the disk 
      and hence is proportional to $m_{00}$ as expected from the dual 
      geometry ${\cal M}$
       (where the size of the 2-sphere was also proportional to 
      $m$). In ${\cal W}$ 
      the same string will be given by a D4 brane on the vanishing 3-sphere.

In the same way we can T-dualize any singularity that can be represented as
a toric variety into a generalized box of NS branes, with a certain amount
of diamonds frozen.

\end{itemize}

\subsection{Going to the interval: the conifold transition}

We can derive a second T-dual triple of geometry 
T-dual brane setup and mirror geometry by studying
T$_U$ and T$_V$ on the interval theory. Note that
the interval theory can be directly obtained from the box
by brane motions.
First we move all the NS and NS' branes
on top of each other, setting all $m_{ij}$ 
to zero, closing all the boxes and diamonds. This is the conifold point. Now
we see that we have the choice to open up the intervals, by turning
on the $w_i$ and $w_j'$. 

We can follow this transition in the geometry as well.
Let us see what it does to ${\cal M}$. For one we have 
shrunk all
the 2-spheres to zero size, putting us at the most singular point of the
geometry. In addition we have put all the B-fields to zero. So we 
are really sitting at the real codimension 2 locus of K\"ahler moduli
space, where the closed string CFT description goes bad \cite{phases}. This
is once more 
the conifold point. From there we can deform the singularity by
turning on 3-spheres to obtain ${\cal M}_T$
and this is precisely what corresponds to
turning on the $w_i$ and $w_j'$ in the brane picture. 
This is a (non-abelian) conifold transition \cite{vafaconi}.
We went from the blowup of the orbifolded
conifold ${\cal C}_{kl}$ to its deformation. 
Let us see that T$_V$ still works. The
$w_i$, $w'_i$ must now be 
       identified with complex structure deformations.
The geometry has to have a ${\mathbb C}^* \times {\mathbb C}^*$ 
fibration which degenerates
over those points. This leads us to
    $$xy = \prod_{i=1}^k (z-w_i)$$
       $$uv = \prod_{j=1}^l (z-w'_i)$$
as the T-dual geometry, indeed.

Last but not least we can study the effect on ${\cal W}$. 
In going to ${\cal W}_T$, the mirror of ${\cal M}_T$, we this
time send all the 3-spheres to zero size and then turn on
blowup modes, taking us from the deformed generalized conifold
${\cal G}_{kl}$ to its blowup.

%%%%%%%%%%%%%%%%%%%%%%%%%%%%%%%%%%%%%%%%%%%%%%%%%%%%%%%%%%%%%
%       SECTION:  PROBING THE MIRROR GEOMETRIES
%%%%%%%%%%%%%%%%%%%%%%%%%%%%%%%%%%%%%%%%%%%%%%%%%%%%%%%%%%%%%

\section{Probing the mirror geometries}
\label{sectionprobing}

\subsection{Introducing the probe: elliptical models}

As a next step we want to introduce $M$ D3 brane probes on top of
our geometry. This way we break the supersymmetry down to 4
supercharges and get interesting $N=1$ 4d gauge theory.
The deformation parameters of the singularity appear
as parameters in the gauge theory, the moduli space of
the gauge theory describes the motion of $M$ D3 branes
on the singular space. These probe theories have
received a lot of attention recently. They give rise to conformal field
theories and have a dual $AdS$ description.

In principle we could take any of the four geometries we introduced, compactify
type IIB on it and then put a D3 brane probe on top of the singularity.
The two situations we are going to study
are $M$ D3 branes on the blowup of the generalized conifold ${\cal G}_{kl}$
 (on ${\cal W}_T$)
and $M$ D3 branes on the blowup of the orbifolded conifold
${\cal C}_{kl}$ (on ${\cal M}$).

Performing our two T-dualities $T_U$ and $T_V$ we will find two different
realizations of each of the probe theories. The background geometry will
transform precisely as we discussed in the last section. This way
\begin{itemize}
\item $M$ D3 brane probes of the blowup of the generalized conifold
${\cal W}_T$ are $T_U$ dual to D4 branes on an interval defined
by $w_i$ and $w_j'$ and $T_{mirror}$
to D6 branes wrapping 3-cycles in ${\cal M}_T$
\item $M$ D3 brane probes of the blowup of the orbifolded conifold
${\cal M}$ are $T_V$ dual to D5 branes on a box defined
by $m_{ij}$ and $T_{mirror}$
to D6 branes wrapping 3-cycles in ${\cal W}$.
\end{itemize}

We will have to deal with what is usually referred to
as elliptical models in the literature \cite{witten4d,haur}.
That is the 6 direction of the interval or the 48 direction of the
box are actually compact, leaving no room for semi-infinite branes.
All D-brane groups will actually be gauged. 

\subsection{The generalized conifold and the interval}

First we would like to consider the gauge theory
on the world volume of $M$ D3 brane probes on
the blowup of a generalized conifold singularity\footnote{Similar
setups have been discussed recently in \cite{raduconi}.}.
This gauge theory is given e.g. in \cite{Uranga} and
can be read off most easily in the dual brane setup we are about to
describe.
In the last section we showed that
this geometry is $T_U$ dual to NS and NS' branes on a circle,
forming intervals
with 67 separations given by $w_i$ and $w_j'$, all the $m_{ij}$ being
zero. As utilized in \cite{Uranga,dasguptamukhi}
this means that the $M$ D3 brane probes turn into an elliptical model
with $M$ D4 branes wrapping the circle. It is straight forward
to read off the gauge theory from this according to the standard
HW rules. Of course it agrees perfectly with the one obtained
from applying a standard orbifold procedure directly on the
conifold gauge theory of \cite{klebwit}.

There is yet another realization of the same gauge theory.
Performing the whole $T_{mirror}=T_{468}$ we can turn
${\cal W}_T$, the blowup of the generalized conifold on 
which we originally put the D3 brane probes, into ${\cal M}_T$,
the deformation of the orbifolded conifold.
The theory with which we have to compare is that on the mirror
of the D3 probe, that is a D6 brane wrapping SUSY 3-cycles in ${\cal M}_T$.
But this is precisely the situation discussed in \cite{ooguri}.
The parameters $w_i$ and $w_i'$ in ${\cal M}_T$, given by
(\ref{zkxkpd}) determine the loci in the $z$
plane where the $\mathbb{C}^* \times \mathbb{C}^*$
fibration degenerates. As found in \cite{ooguri} in order
to have a BPS state the $w_i$ and $w_i'$ have to align along
a line in the $z$ plane. Since the $S^1 \times S^1$
fibration degenerates over
$w_i$ and $w_i'$, we can regard this fibration over the interval between
neighboring $w_i$ and $w_i'$ as a 3-cycle. In \cite{ooguri}
it was shown that this 3-cycle is $S^3$ and $S^2 \times S^1$ respectively,
depending on whether neighboring points are a $w$, $w'$ pair or both
$w$ (both $w'$). In the former case one obtains a quartic superpotential,
in the latter case an $N=2$ like setup. Obviously this
yields the same gauge theory as the D3 probe on ${\cal W}_T$ and the D4 brane
on the interval.

\subsection{D5 branes on the box: the modified box rules}

The second theory we would like to consider are $M$ D3 branes on
an $\mathbb{Z}_k \times \mathbb{Z}_l$ orbifolded conifold. 
As shown above, the geometry dualizes under $T_V$ into
brane boxes where the NS5 brane skeleton wraps the
curve $\sum_{i,j=0}^{k,l} m_{ij} z^i w^j$. $k+l-2$ of the $m_{ij}$
parameters can be associated to brane positions, while
the other $kl$ parameters correspond to diamonds, that
is the hypermultiplets sitting at the NS NS' intersections, whose
vev smoothes out the singular intersections.

The probe D3 branes turn into D5 branes living on these boxes and diamonds.
Again this should in principle be a very useful duality in the sense that
we can read off the associated gauge theories by using some analogue
of the HW rules. In addition some information about the corresponding
quantum gauge theory should be obtainable by lifting the setup to M-theory.

\begin{figure}[htb]
  \refstepcounter{figure}
  \label{figurecondia}
  \begin{center}
  \makebox[8cm]
  {
     \begin{turn}{0}%
        \epsfxsize=8cm
        \epsfysize=7cm
        \epsfbox{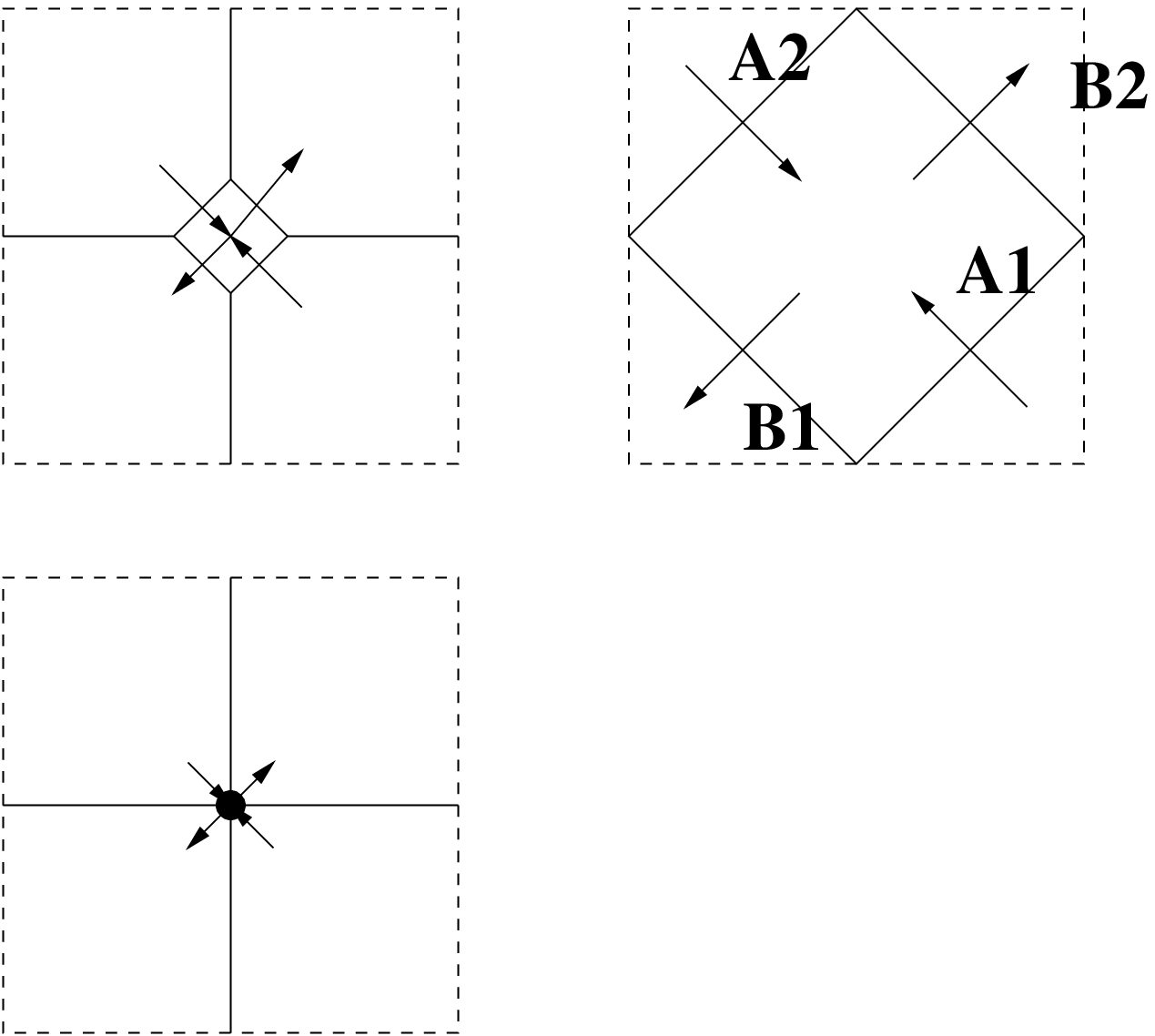}
     \end{turn}
  }
  \end{center}
  \vbox{
    \hbox{\hspace{2cm}{\bf Fig.\thefigure.} Upper left~~:  the box 
          with generic B-value;
         }
    \hbox{\hspace{3.2cm} Upper right: maximal B-value;} 
    \hbox{\hspace{3.2cm} Lower left~~:  Taking B to 0 sending one}
    \hbox{\hspace{5.6cm} gauge coupling to infinity.} 
  }
\end{figure}

\noindent
In order to understand our rules it is best to start with the easiest
example, the conifold ${\cal C}$,
eq.(\ref{con}), itself. The dual description just is that of a 
single NS and NS' brane on a square torus, as depicted in the upper left 
corner of Fig.\ref{figurecondia}.
%%%%%%%%%%%%%%%%%%%%%%%%%%%%%%%%%%%%%%%%%%%%%%%%%%%%%
%   BILD: THE CONIFOLD ITSELF
%%%%%%%%%%%%%%%%%%%%%%%%%%%%%%%%%%%%%%%%%%%%%%%%%%%%%
%%%%%%%%%%%%%%%%%%%%%%%%%%%%%%%%%%%%%%%%%%%%%%%%%%%%%
%
%%%%%%%%%%%%%%%%%%%%%%%%%%%%%%%%%%%%%%%%%%%%%%%%%%%%%
\noindent
The conifold has one blowup parameter, corresponding
to the one diamond sitting at the intersection.
As long as we keep the size of the 2-sphere zero, the B-flux
through the sphere will correspond to the size of the diamond.
As we have argued in the last section, the curve describing the
diamond actually supports a non-trivial $S^1$ on
which the D5 brane can end, so the gauge
theory will have two group factors, $SU(M) \times SU(M)$. The
inverse gauge couplings are proportional to the area of
the corresponding faces.
\noindent
There is a special point, when the diamond has the same area as the
other gauge group, that is the diamond occupies half of the torus.
In this case we know that we have to recover the standard
conifold gauge theory of \cite{klebwit}. 
This can easily be implemented using the simple brane rules
specified in the upper
%%%%%%%%%%%%%%%%%%%%%%%%%%%%%%%%%%%%%%%%%%%%%%%%%%%%%
%   BILD: URANGA
%%%%%%%%%%%%%%%%%%%%%%%%%%%%%%%%%%%%%%%%%%%%%%%%%%%%%
\begin{figure}[htb]
  \refstepcounter{figure}
  \label{figureUranga}
  \begin{center}
  \makebox[7cm]
  {
     \begin{turn}{0}%
        \epsfxsize=6.5cm
        \epsfysize=6.5cm
        \epsfbox{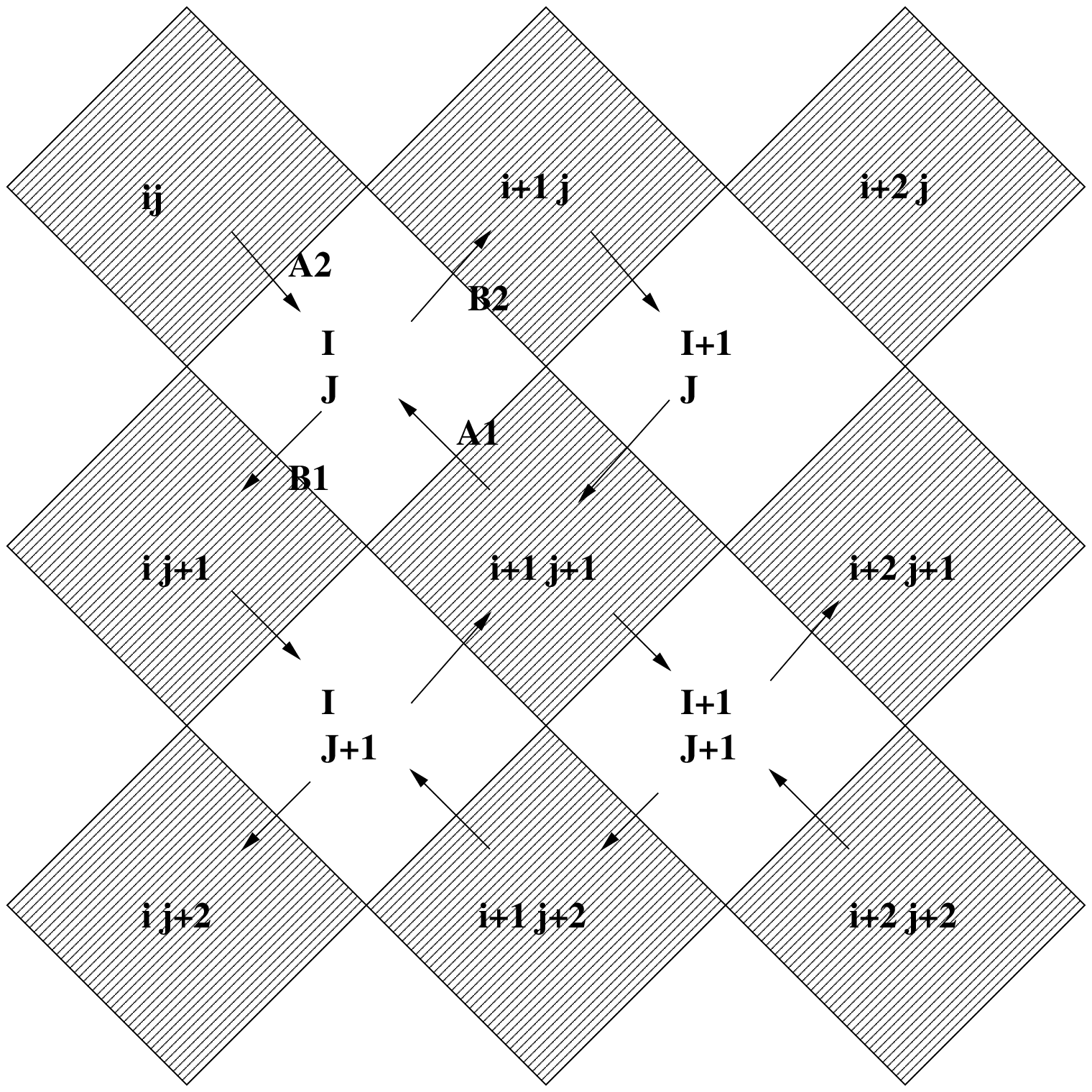}
     \end{turn}
  }
  \end{center}
  \begin{center}
    {\bf Fig.\thefigure}{\bf.} The diamond rules at the point
                of maximal B-fields.
  \end{center}
\end{figure}
%%%%%%%%%%%%%%%%%%%%%%%%%%%%%%%%%%%%%%%%%%%%%%%%%%%%%
%
%%%%%%%%%%%%%%%%%%%%%%%%%%%%%%%%%%%%%%%%%%%%%%%%%%%%%
\noindent
right corner. 
We have to demand, that half of the matter multiplets
we would naively expect are projected out. 
The orientation of the arrows seems quite arbitrary. Indeed we will
see that the orientation can be changed and that this
corresponds to performing flop transitions in the dual
geometry.
Indeed one can easily establish that these rules also are capable
of realizing more complicated setups. Generically,
the gauge theory on the $\mathbb{Z}_k \times \mathbb{Z}_l$ orbifolded
conifold has a \hbox{$SU(M)^{kl} \times SU(M)^{kl}$} gauge group. In
our picture the gauge group factors will correspond to the $kl$ diamonds and
the $kl$ boxes respectively. Again it is easiest to compare at
the point, where all gauge couplings are equal. In this case, both
the diamonds as well as the boxes degenerate to rhombes, as
pictured in Fig.\ref{figureUranga}, where we denoted them as filled and unfilled
boxes. Generalizing our $A$ and $B$ fields from above we will find
that the matter fields transform as (where the two sets of
$kl$ gauge groups are indexed by small and capital letters respectively)

\begin{eqnarray*}
  \begin{array}{ll}
(A_1)_{i+1,j+1;I,J} & (\fund_{i+1,j+1}, \bar{\fund}_{I,J}) \\
(A_2)_{i,j;I,J} &    (\fund_{i,j}, \bar{\fund}_{I,J}) \\
(B_1)_{I,J;i,j+1}&   (\bar{\fund}_{i,j+1}, \fund_{I,J}) \\
(B_2)_{I,J;i+1,j}&   (\bar{\fund}_{i+1,j}, \fund_{I,J})
   \end{array}
\end{eqnarray*}

which are exactly the rules expected \cite{Uranga}. This proposal can
also easily deal with the situation of non-trivial
identifications on the torus as discussed in \cite{haur}.
In addition there will be quartic superpotential for every
closed rectangle, the relative sign being given by the orientation

\begin{eqnarray*}
W&=&\sum_{i,j} (A_1)_{i+1,j+1;I,J} (B_1)_{I,J;i,j+1}
(A_2)_{i,j+1;I,J+1} (B_2)_{I,J+1;i+1,j+1} - \\
&&\sum_{i,j} (A_1)_{i+1,j+1;I,J} (B_2)_{I,J;i+1,j}
(A_2)_{i+1,j;I+1,J} (B_2)_{I+1,J;i+1,j+1} 
\end{eqnarray*}

We do not expect that this picture changes when we take the sizes
of box and diamond to differ.
We will still see the $A$ and $B$ fields.
Only the relative couplings will change and
no new fields or interactions appear, since
they certainly don't in the dual geometry. The singular conifold points
 correspond
to the situations where diamonds close. From the field theory
point of view this just means that we take the corresponding gauge
coupling to infinity. As in the standard HW situation with only
parallel NS branes this corresponds to a strong coupling
fixed point with possibly enhanced global symmetry if several NS branes
coincide.

Another interesting question to consider is to ask ourselves what happens
when we blow up the spheres to finite size. This now should correspond
to some mode of the diamond that ``rotates'' it away out of the 48 plane into
the 59 plane. According to common lore this should correspond to
a FI
term in the gauge theory. We will no longer be able to
support a D5 brane stretched inside the diamonds in a supersymmetric fashion,
independent of their
size (that is the B-field)\footnote{This is very similar to what happens
on the interval: blowing up a sphere corresponds to moving off
an NS brane in the 7 direction. Since in order to preserve
supersymmetry branes are only allowed to stretch along the 6
direction this effectively reduces the
number of gauge groups (the number of intervals) by one. 
The 6 position of the brane we moved away (the B-field on the blown
up sphere) does not affect the massless matter content anymore.}.
Since we expect that the center of mass $U(1)$s are frozen out
as in \cite{witten4d}, the FI term will be reinterpreted as usual
as a baryonic branch. Especially there should exist a baryonic 
branch along which we reduce to
the orbifold gauge theory.

Indeed as shown in \cite{Uranga} the gauge theories we described here
do have such a baryonic branch. Giving a vev to (say) all the $A_2$
fields will break each $SU(M)_{ij} \times SU(M)_{IJ}$ pair
down to its diagonal $SU(M)_{ab}$ subgroup. The remaining massless
fields after the Higgs mechanism are
\begin{eqnarray*}
  \begin{array}{lcll}
D_{a+1,b+1;a,b}&=&(A_1)_{a+1,b+1;A,B} & (\fund_{a+1,b+1}, \bar{\fund}_{a,b}) \\
H_{a,b;a,b+1}&=&(B_1)_{A,B;a,b+1}&   (\fund_{A,B}, \bar{\fund}_{a,b+1}) \\
V_{a,b;a+1,b} &=&(B_2)_{A,B;a+1,b}&   (\fund_{A,B}, \bar{\fund}_{a+1,b}) \\
   \end{array}
\end{eqnarray*}
with the remaining superpotential:
\begin{eqnarray*}
W&\sim&\sum_{a,b} D_{a+1,b+1;a,b} H_{a,b;a,b+1}
V_{a,b+1;a+1,b+1} - \\
&&\sum_{a,b} D_{a+1,b+1;a,b} V_{a,b;a+1,j}
H_{a+1,b;a+1,b+1}
\end{eqnarray*}
which are precisely the box rules of \cite{hanzaf}, as claimed.
Note that the diagonal $D$ fields are not special at all, they arise just
from the fundamental $A$, $B$ degrees of freedom of the generalized box.

A small complication arises once we
%%%%%%%%%%%%%%%%%%%%%%%%%%%%%%%%%%%%%%%%%%%%%%%%%%%%%
%   BILD: DIAMONDS
%%%%%%%%%%%%%%%%%%%%%%%%%%%%%%%%%%%%%%%%%%%%%%%%%%%%%
\begin{figure}[htb]
  \refstepcounter{figure}
  \label{figureorientation}
  \begin{center}
  \makebox[7cm]
  {
     \begin{turn}{0}%
        \epsfxsize=7cm
        \epsfysize=8cm
        \epsfbox{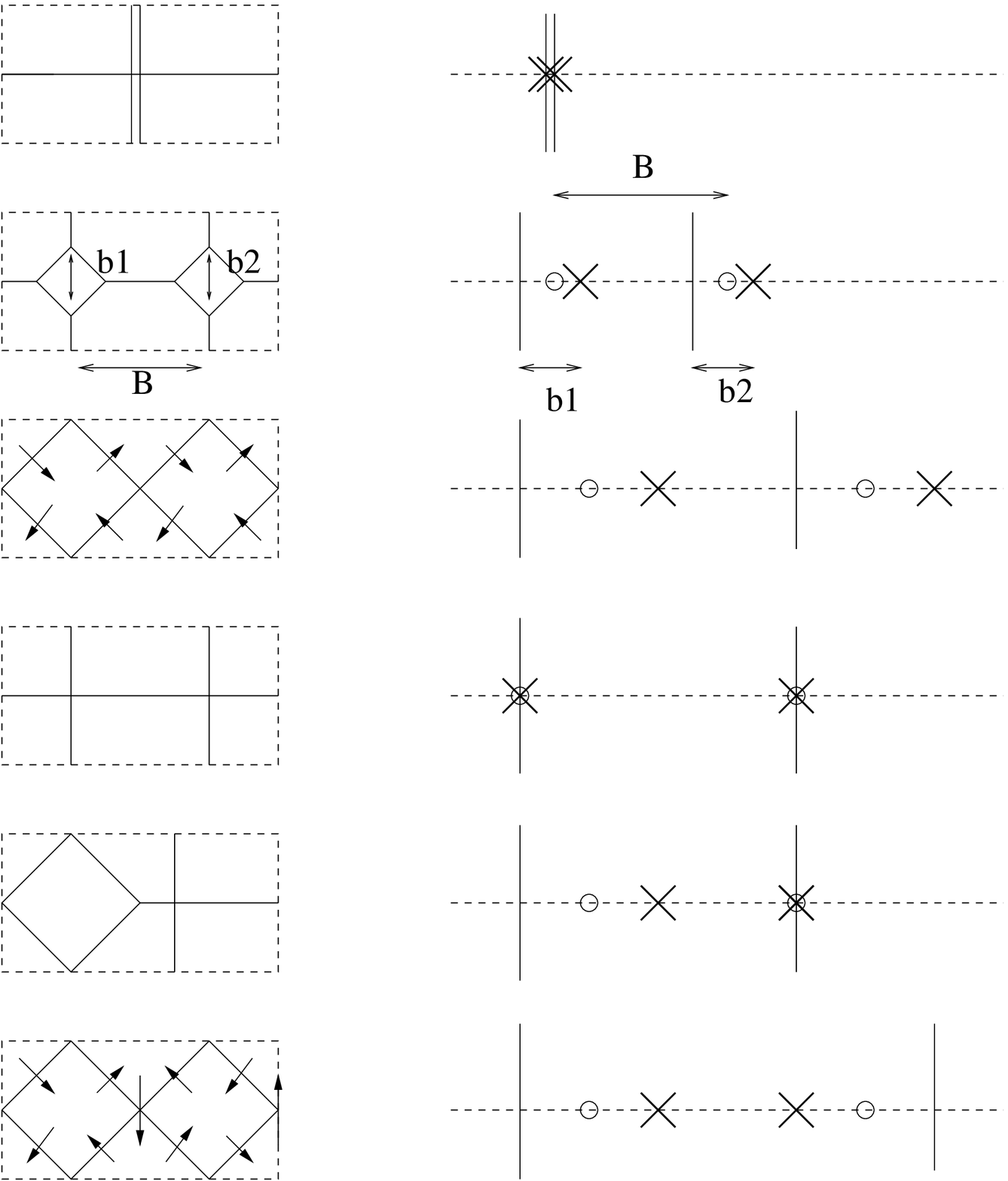}
     \end{turn}
  }
  \end{center}
  \begin{center}
    {\bf Fig.\thefigure}{\bf.} Diamonds do have an orientation.
  \end{center}
\end{figure}
%%%%%%%%%%%%%%%%%%%%%%%%%%%%%%%%%%%%%%%%%%%%%%%%%%%%%
%
%%%%%%%%%%%%%%%%%%%%%%%%%%%%%%%%%%%%%%%%%%%%%%%%%%%%%
consider situations that
are more involved than the conifold. For simplicity
let us study the case of the ${\mathbb{Z}}_2$ orbifolded conifold. Since
this can as well be thought of as the ${\cal G}_{22}$ generalized conifold, 
it has an interval
dual as well as a box dual. Both of them are displayed in 
Fig.\ref{figureorientation} for various values of the B-fields.
The gauge group is $SU(M)^4$. We should see 3 B-fields governing
the relative sizes of the gauge couplings. According to our scenario
this will correspond to one relative brane position $B$ and the
sizes of two diamonds $b_1$ and $b_2$. In the interval picture
$b_{1,2}$ will be the distance
between NS$_{1,2}$ and NS'$_{1,2}$ while $B$ is the distance
between the center of masses of the two NS NS' pairs, denoted
as circles in Fig.\ref{figureorientation}.
Take the circle to have circumference 2 and the torus to have 
sides 2 and 1. Since B-fields (=inverse gauge couplings) 
are length on the interval and areas on the torus, 
in these units the area of a given 
gauge group on the torus should have the same numerical value as the 
corresponding length on the circle (total area=total length=2).
The third picture in Fig.\ref{figureorientation} shows $B=1$ $b_1=b_2=1/2$
Both sides have 4 gauge groups of size $1/2$.

It is easy to identify
in both theories the point where all gauge couplings are equal, the
point where all B-fields are zero (the most singular point) and
the point where the setup looks like two separated conifolds.
Similarly for all positive values of the $b_i$ and of $B$
we can read off the gauge theory from the diamonds, just using the
standard $A$ and $B$ fields, representing the diamonds as rhombes of
area $b_i$.
However from the interval it is clear, that we can also pass
an NS' brane through an NS brane, performing Seiberg duality
on the gauge theory and simultaneously changing the sign
of one of the $b_i$ fields \cite{Uranga,unge}.
If we set $b_1=b_2=-1/2$ the picture looks the same as for $b_1=b_2=1/2$.
The overall sign does not matter. However the sixth picture of 
Fig.\ref{figureorientation}
shows a setup where the signs of the $b_i$ differ. We should
assign our diamonds an orientation in order to be able to
address this issue. This orientation assigns whether
the $A$ or the $B$ fields point outward or inward, the other
doing the opposite. The rules we have introduced are valid
for the case that all orientations are equal. The situation
with opposite orientations is slightly more complicated. The rules
can be determined by comparing with the interval. Whenever
the arrows point around the closed rectangle we write down
a quartic superpotential. If diamonds with different orientation
touch, we will have to introduce additional `meson' fields
with cubic superpotential (see the 6th picture
in Fig.\ref{figureorientation}). Since this
inversion of orientation should correspond to Seiberg duality
in the field theory, we basically found this way
a realization of $N=1$ dualities in the box and diamond picture!
It would be clearly interesting to
pursue this point further, for example by
studying theories with orientifolds. This may give us a hint of a
brane realization of Pouliot like dualities \cite{pouliot} and spinors, since
it is easy to realize the magnetic side of these theories
in the box and diamond picture using orientifolds.

Last but not least we should be able to see the same gauge groups in
the third T-dual realization as well, that is from D6 branes wrapping
the 3-cycles of the deformed generalized conifold geometry (\ref{conid})
$$xy= \sum_{i,j=1}^{k,l} m_{ij} u^i v^j$$
in the same spirit as above following \cite{ooguri}. It would
be interesting to work this out and see if some properties of the
gauge theory can be better understood in this language.

%%%%%%%%%%%%%%%%%%%%%%%%%%%%%%%%%%%%%%%%%%%%%%%%%%%%%%%%%%%%%
%       SECTION:  GEOMETRIZING EVERYTHING
%%%%%%%%%%%%%%%%%%%%%%%%%%%%%%%%%%%%%%%%%%%%%%%%%%%%%%%%%%%%%

\section{Mirror branes and Domain Walls}
\label{sectioneverything}

\subsection{The mirror branes}

The D3 brane
probe we have been considering so far maps to a D4 brane
on the interval and a D5 brane in the box respectively. We identified
the corresponding gauge theories above. For a special
subclass of models we were considering
we can actually perform both. These geometries
are those whose toric diagram is given by two rows of $k$ points. Viewing
them as ${\mathbb{Z}}_k$ 
orbifolded conifolds ${\cal C}_k$, they (or better their blowup)
turn into a box with 1 NS' and $k$ NS under T$_V$. 
We can as well describe them as a ${\cal G}_{kk}$ generalized 
conifold and hence T$_U$ dualize them into an interval with $k$ NS and $k$ NS'
branes. 
According to our philosophy
these two ways of realizing the gauge theory should actually be mirror
to each other! We turned one HW setup into its `mirror branes'.

Now we can try to solve these gauge theories via the lift to M-theory.
Interestingly enough, the intervals lift via SUSY 2-cycles in $\mathbb{R}^6$
while the boxes lift via SUSY 3-cycles \cite{miemiecetal}
in $\mathbb{R}^6$. So for every 3-cycle
we should find a dual 2-cycle encoding the same information and vice versa.

\subsection{Putting together intervals and boxes}
Above we obtained an $N=1$ $d=4$ gauge theory from
intervals in type IIA and boxes in type IIB setups respectively. 
Of course we can as well 
build a box in type IIA or an interval in type IIB in order to
obtain odd dimensional gauge theories with 4 supercharges.
The singular point should correspond to having all NS branes coinciding.

We can do both together, that is put branes
on the box and the interval simultaneously, provided
we put in enough NS branes so that we can open up both, a box
and an interval. From the dual geometry point of view this
corresponds to consider manifolds with both complex and K\"ahler
deformations turned on simultaneously. An interesting example
is type IIA with
NS 012345, NS' 012389, D4 01236, D4 01248. It is easy
to convince oneself, that this now lifts to M-theory via a SUSY 3-cycle
in $G_2$. That is we now break another half of the SUSY, leaving us
with 2 unbroken supercharges, or $N=1$ in $d=3$. Note
that this gauge theory actually only lives on the boxes, since
the interval theory is 4d while the box theory is 3d.
Things become more interesting if we compactify the $x^3$ direction. In this
case both the interval and the box give 3d gauge theories. 

These brane setups fit nicely into the framework of brane cubes. These
also lead to 2 supercharges. They lift via 
$G_2$ and $SU(4)$ 4-cycle respectively
and are dual to probes on $SU(4)$ and $G_2$ orbifold. Now we have a 3rd kind
of brane setup in this league, which lifts via $G_2$ 3-cycle and should
probably also be dual to probes on a $G_2$ singularity.

Note that from the point of view of the four dimensional theory 
on the D4 branes
on the interval, the D4 branes on the box look like domain walls
(they are localized in $x^3$). This is nice, since Witten argued 
\cite{witteng2}
before that domain walls in $d=4$, $N=1$ gauge theory should
be associated to M5 branes on $G_2$ 3-cycles.

\section{Summary}
\label{sectionsummary}

Let us briefly summarize the main results of the paper.
For two classes of non-compact
(complex 3-dimensional) Calabi-Yau spaces
we constructed  the T-dual NS brane configurations.
Specifically blowups (resp. deformations)
of orbifolded conifold singularities,
denoted by ${\cal C}_{kl}$, are $T_V$ dual to boxes (resp. intervals) of NS
branes, whereas blowups (resp. deformations) of generalized conifold
singularities, called ${\cal G}_{kl}$, are $T_U$ dual to intervals
(resp. boxes)
of NS branes. Since the composition of $T_U$ and $T_V$ corresponds to a
T-duality with respect to three isometrical $U(1)$ directions of ${\cal M}$
resp. ${\cal W}$, it should not come as a surprise that ${\cal C}_{kl}$ and 
${\cal G}_{kl}$ are actually
mirror pairs.
The K\"ahler resp. complex structure parameters
of the geometric singularities correspond to positions 
of the NS branes in the dual brane picture. Moreover the conifold
transition for the non-compact Calabi-Yau  spaces ${\cal C}_{kl}$ or
${\cal G}_{kl}$
via shrinking \hbox{2-cycles} and growing up \hbox{3-cycles}  precisely
corresponds to the transition between the box and interval theory or
vice versa, by first moving all NS branes on top of each other and then
removing them into different directions.
All this is summarized in Fig.\ref{figureSummary} below. 
%%%%%%%%%%%%%%%%%%%%%%%%%%%%%%%%%%%%%%%%%%%%%%%%%%%%%
%   BILD: THE PROPOSED PICTURE
%%%%%%%%%%%%%%%%%%%%%%%%%%%%%%%%%%%%%%%%%%%%%%%%%%%%%
\begin{figure}[htb]
  \refstepcounter{figure}
  \label{figureSummary}
  \begin{center}
  \makebox[10cm]
  {
     \begin{turn}{0}%
        \epsfxsize=9cm
        \epsfysize=7cm
        \epsfbox{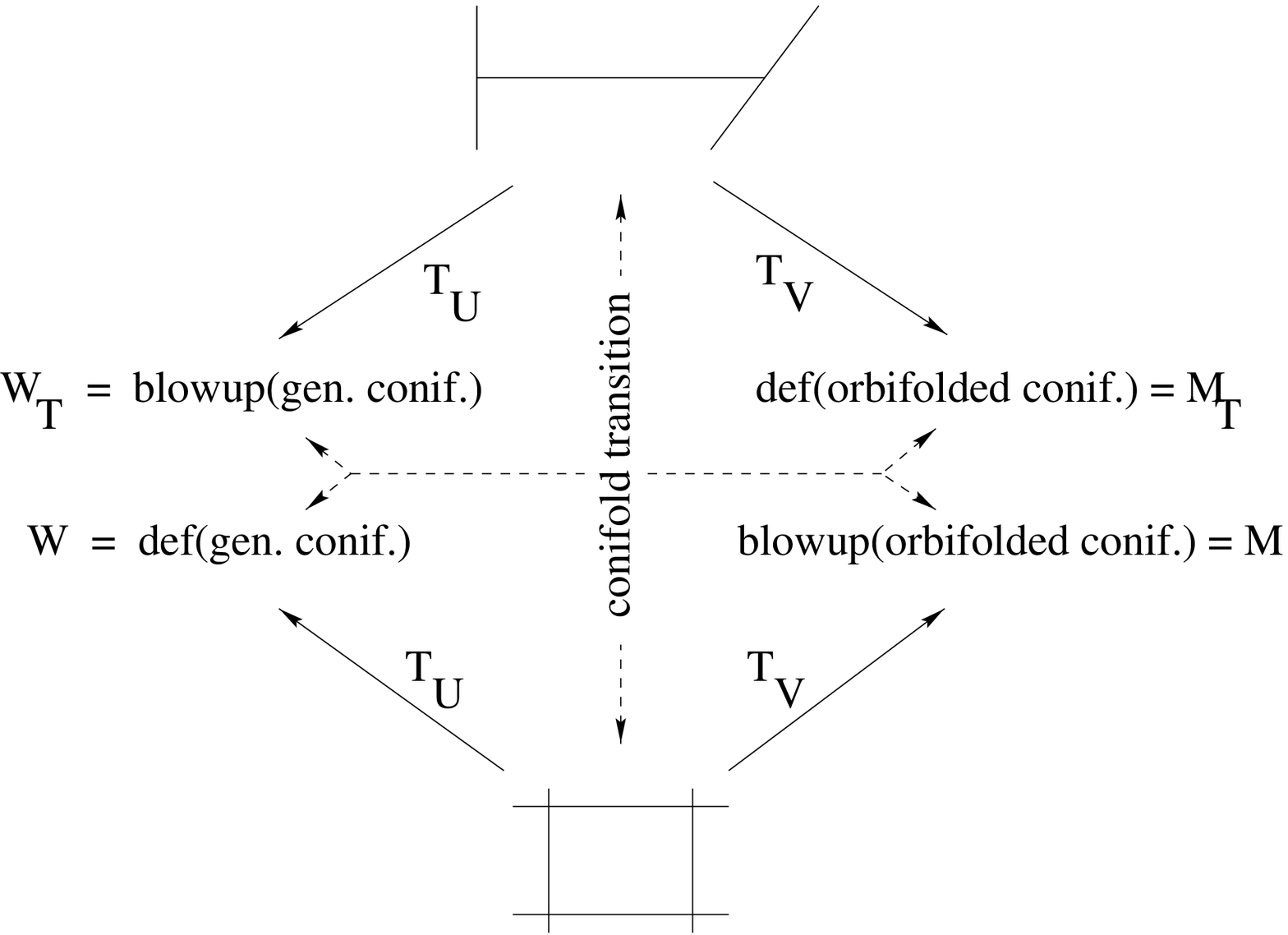}
     \end{turn}
  }
  \end{center}
  \center{{\bf Fig.\thefigure}{\bf.} The proposed picture.}
\end{figure}
%%%%%%%%%%%%%%%%%%%%%%%%%%%%%%%%%%%%%%%%%%%%%%%%%%%%%
%
%%%%%%%%%%%%%%%%%%%%%%%%%%%%%%%%%%%%%%%%%%%%%%%%%%%%%
\noindent

Constructing gauge theories from branes,
the geometric singularities as well as the NS brane configurations
serve as backgrounds, which are probed by a certain number of D branes.
We have seen that the ``mirror map'' does not change the corresponding
gauge theories. At the conifold point some of the gauge couplings go 
to infinity.

In order to establish the
duality between conifold singularities and brane boxes we had to
generalize the concept of brane boxes by also including brane diamonds.
We
formulate rules for deriving the matter content of the
gauge theories living on boxes and diamonds. 
Along a baryonic branch of the gauge theory, which corresponds to
partially resolving the conifolds ${\cal C}_{kl}$
to the orbifold singularities ${\cal O}_{kl}$
we recover the orbifold gauge theories from our general rules.

Blowups (or deformations) of
certain geometries, namely ${\cal C}_{1k}\equiv {\cal G}_{kk}$,
allow both for a dual brane box as well as for a dual interval
description. It follows that the corresponding gauge theory on the
interval and on the brane box are mirror to each other.
This observation could be useful for the investigation of the 
non-perturbative
quantum dynamics of these kind of $N=1$ gauge theories: namely for every
supersymmetric 2-cycle which describes the dynamics of the interval
theory embedded in M-theory, there should exist a mirror supersymmetric
3-cycle for the brane box theory also embedded in M-theory.
It would be interesting to work out this mirror map between
2- and 3-cycles explicitly.
Moreover one could expect that due to quantum corrections the physics
of the gauge theories at the
conifold point is not as singular as in the classical description
we have discussed.
Finally, it would be also interesting to relate the brane constructions
of $N=1$ supersymmetric gauge theories, considered here, to
the geometric engineering approach, where various branes are 
wrapped around
non-trivial cycles of Calabi-Yau 4-folds or manifolds of $G_2$ holonomy.

\section*{Acknowledgments}

We would like to thank H. Skarke and
especially A. Hanany, P. Mayr and M. Strassler for very useful discussions.
The work of M.A and A.K. is supported in part by funds provided
by the U.S. Department of Energy (D.O.E.) under cooperative
research agreements \# DE-FG03-92-ER40701 and \# DF-FC02-94ER40818
respectively.
The work of A.M. and D.L. is partially supported by the E.C. project
ERBFMRXCT960090 and by the Deutsche Forschungs Gemeinschaft.

%% \CharacterTable
%%  {Upper-case    \A\B\C\D\E\F\G\H\I\J\K\L\M\N\O\P\Q\R\S\T\U\V\W\X\Y\Z
%%   Lower-case    \a\b\c\d\e\f\g\h\i\j\k\l\m\n\o\p\q\r\s\t\u\v\w\x\y\z
%%   Digits        \0\1\2\3\4\5\6\7\8\9
%%   Exclamation   \!     Double quote  \"     Hash (number) \#
%%   Dollar        \$     Percent       \%     Ampersand     \&
%%   Acute accent  \'     Left paren    \(     Right paren   \)
%%   Asterisk      \*     Plus          \+     Comma         \,
%%   Minus         \-     Point         \.     Solidus       \/
%%   Colon         \:     Semicolon     \;     Less than     \<
%%   Equals        \=     Greater than  \>     Question mark \?
%%   Commercial at \@     Left bracket  \[     Backslash     \\
%%   Right bracket \]     Circumflex    \^     Underscore    \_
%%   Grave accent  \`     Left brace    \{     Vertical bar  \|
%%   Right brace   \}     Tilde         \~}

\bibliography{diamonds}
\bibliographystyle{utphys}

\end{document}